\documentclass{article}

    \PassOptionsToPackage{numbers, compress}{natbib}
\usepackage[preprint]{neurips_2026}

\usepackage[utf8]{inputenc} % allow utf-8 input
\usepackage[T1]{fontenc}    % use 8-bit T1 fonts
\usepackage{hyperref}       % hyperlinks
\usepackage{url}            % simple URL typesetting
\usepackage{booktabs}       % professional-quality tables
\usepackage{amsfonts}       % blackboard math symbols
\usepackage{nicefrac}       % compact symbols for 1/2, etc.
\usepackage{microtype}      % microtypography
\usepackage[table]{xcolor}         % colors
\usepackage{amsmath}
\usepackage{pifont}
\usepackage{graphicx} 
\usepackage{wrapfig}
\usepackage{listings}
\usepackage[most]{tcolorbox}
\usepackage{enumitem}
\newcommand{\zshotbest}[1]{#1\rlap{\hspace{0.1pt}\textsuperscript{\ding{72}}}}

\title{ProCrit: Self-Elicited Multi-Perspective Reasoning with Critic-Guided Revision for Multimodal Sarcasm Detection}
\author{%
  Yingjia Xu$^{1,3}$\quad
  Jiulong Wu$^{2}$\quad
  Bowen Zhang$^{1}$\quad
  Baokui Guo$^{3}$\quad
  Siyuan Chai$^{3}$\quad
  Min Cao$^{1}$\thanks{Corresponding author.}\\
    $^{1}$School of Computer Science and Technology, Soochow University, Suzhou, China\\
  $^{2}$Baidu Inc., Beijing, China\\
  $^{3}$Zhipu AI, Beijing, China\\
  \texttt{\{yjxuxuyj, mcao\}@suda.edu.cn}
}

\begin{document}

\maketitle

\begin{abstract}
% Multimodal sarcasm detection requires identifying incongruities between literal expression and intended meaning, often across image and text.
% Because sarcasm can arise through diverse mechanisms, such as verbal irony, hyperbole, rhetorical questions, and cultural allusions, different samples require different analytical perspectives and different ways of combining them.
Multimodal sarcasm detection requires reasoning over cross-modal incongruities between literal expression and intended meaning, yet the specific analytical perspectives needed vary across samples due to the diversity of sarcastic mechanisms.
While recent methods make this analytical process explicit, they still rely on fixed, predefined perspectives that operate independently under hand-crafted routing rules.
We argue that multimodal sarcasm detection instead calls for \emph{self-elicited multi-perspective reasoning}, where a model autonomously generates the perspectives needed for each sample and progressively integrates them into a coherent analysis.
To realize this goal, we propose \textbf{ProCrit}, a \textbf{Pro}posal--\textbf{Crit}ic two-agent framework with a proposal agent for multi-perspective reasoning and a critic agent for external evaluation and targeted revision guidance.
First, to overcome the lack of process-level supervision in existing sarcasm datasets, ProCrit synthesizes process-level reasoning annotations through a \emph{dynamic-role agentic rollout}: a strong vision-language model sequentially spawns analytical roles within a shared context, and the resulting multi-role trajectories are flattened into sequences that preserve cross-perspective dependencies while enabling efficient autoregressive generation.
Second, to improve reasoning reliability, ProCrit adopts a \emph{draft--critique--revise} paradigm in which an independent critic identifies reasoning deficiencies and provides targeted natural-language feedback for directed revision.
Finally, we develop a \emph{mutual-refinement} training framework that jointly optimizes proposal drafting and feedback-guided revision via dual-stage reinforcement learning, while refining the critic agent according to the actual effectiveness of its feedback.
Experiments on three widely used benchmarks demonstrate the effectiveness of ProCrit. 
% The code will be made publicly available.
% : the proposal agent drafts an initial reasoning, the critic identifies specific flaws and provides targeted natural-language feedback, and the proposal revises its output guided by the feedback.
\end{abstract}
\begingroup
\renewcommand*{\addcontentsline}[3]{}
%% ================================================================
%%  SECTION 1 — Introduction (placeholder)
%% ================================================================

% 1 — 任务引入
% 定义多模态讽刺检测 → 用具体例子说明跨模态 incongruity → 强调核心挑战是需要 reasoning，且不同样本需要不同的分析角度组合

% 2 — 现有方法局限（层层递进）
% 端到端分类方法 → 黑盒，能判断但不能解释
% LLM/VLM prompting 方法（S3 Agent、Commander-GPT、IRONIC）→ 有中间推理了（只是碎片化的并行输出的独立分析片段，不是真正连贯的、有逻辑依赖的多视角推理。我们的 dynamic-role rollout 才是有跨视角依赖、渐进式构建的连贯推理链。），但三大硬伤：固定视角、无交互、无训练/修正

% 3 — 我们的 process-aware 视角 + 三个设计原则
% 核心论点：模型应该产出可审视的推理链，并且在推理有缺陷时能被修正
% 三原则：动态视角、外部评估（引 Huang et al. 证明自修正不可靠）、训练优化推理能力

% 4 — 方法概述（CRAFT）
% Proposal agent（draft → react）+ Critic agent（score + NL feedback）
% 两个核心组件：dynamic-role agentic rollout 数据合成 + dual-stage GRPO
% Critic 的 actionability reward 也提了

% 5 — 四条 Contributions
% Dynamic-role agentic rollout 数据合成
% Draft–critique–react 推理范式
% Dual-stage GRPO + critic actionability reward
% MMSD2.0 实验验证 + ablation

\section{Introduction}
\label{sec:intro}

% Multimodal sarcasm detection, determining whether an image-text pair conveys sarcastic intent, is a key challenge in multimodal understanding.
Multimodal sarcasm detection, determining whether an image-text pair conveys sarcastic intent, has attracted growing interest due to its importance in content moderation, sentiment analysis, and emotionally aware dialogue systems~\cite{joshi2017automatic,maynard2014cares,abulaish2023survey}.
% Sarcasm typically arises from incongruities between what is literally expressed and what is actually meant, and in the multimodal setting these incongruities often span modalities: a caption reading ``what a perfect day'' paired with a photo of a traffic jam in a downpour.
% Detecting such incongruities is challenging because sarcasm manifests through diverse mechanisms---visual irony, emotional mismatch, rhetorical subversion, cultural allusion---and different samples require different combinations of analytical perspectives to uncover the underlying intent.
% ,yet these approaches compress multi-perspective understanding entirely into implicit feature representations, with no explicit process for examining the input from distinct analytical angles.
% to bridge the gap between literal expression and intended meaning
Sarcasm fundamentally relies on incongruities between literal expression and intended meaning, and in the multimodal setting these incongruities often span modalities, \emph{e.g.,} a caption reading ``what a perfect day'' paired with a photo of a traffic jam in a downpour.
% Sarcasm often emerges from incongruities between literal expression and intended meaning and in multimodal settings, such incongruity often span modalities, e.g., a caption reading ``what a perfect day'' paired with a photo of a traffic jam in a downpour.

Detecting such incongruities is challenging because sarcasm arises from diverse mechanisms---verbal irony, hyperbole, rhetorical questions, cultural allusion, among others. 
Exposing the underlying incongruity requires examining the sample from multiple \emph{analytical perspectives}---distinct angles of inspection such as probing a cultural reference, decoding exaggerated expressions, or evaluating scene plausibility against commonsense expectations, as illustrated in Fig.~\ref{fig:perspective_example}.
% Which perspectives are needed and how they should be combined varies across samples, as different samples exploit different mechanisms and embed signals across modalities in different ways.
Which perspectives are needed and how they should be combined varies across samples, as different instances of sarcasm rely on distinct mechanisms and distribute critical cues across text and image in different ways.

To address this challenge, early works~\cite{cai2019multi,liang2022multi,liang2024fusion} have explored increasingly sophisticated cross-modal fusion strategies through attention, graph-based, and contrastive learning mechanisms, in an effort to capture the diverse signals that underlie sarcastic intent.
% These approaches implicitly encode multiple analytical perspectives within learned representations, but the perspectives remain opaque and entangled---the model decides which to activate and how to combine them without any explicit structure, leaving the reasoning process entirely implicit.
% Recent methods~\cite{wang2025s3agent,zhang2025commandergpt,ramakrishnan2025ironic} introduce hand-crafted analytical roles and routing rules by prompting LLMs or VLMs, explicitly prescribing which perspectives the model should adopt when analyzing a given input.
% Recent methods~\cite{wang2025s3agent,zhang2025commandergpt,ramakrishnan2025ironic} go further by prompting LLMs or VLMs with hand-crafted analytical roles and routing rules, explicitly prescribing which perspectives the model should adopt.
% Recent methods~\cite{wang2025s3agent,zhang2025commandergpt,ramakrishnan2025ironic} make the analytical process explicit by prompting LLMs or VLMs with hand-crafted analytical roles, each analyzing independently from a fixed perspective under predefined routing rules.
% Recent methods~\cite{wang2025s3agent,zhang2025commandergpt,ramakrishnan2025ironic} make the analytical process explicit by prompting LLMs or VLMs with a fixed, predefined set of perspectives, 
Recent methods explicitly model the analytical reasoning process by prompting Large Language Models (LLMs) or Vision–Language Models (VLMs) with a fixed, predefined set of analytical perspectives,
each operating independently under hand-crafted routing rules.
Although promising, these methods overlook that the analytical perspectives needed are inherently sample-specific, and thus the fixed, predefined set can hardly cover the diverse reasoning demands across samples;
and that effective multi-perspective analysis should be progressive:
% each new angle of exploration should be informed by what preceding analyses have revealed.
the exploration of each subsequent perspective should 
% build upon what preceding analyses have revealed.
be informed by insights gleaned from prior analytical steps.
% Determining the right perspectives is an integral part of the reasoning itself, not a prerequisite that can be externally specified, calling for \emph{spontaneous multi-perspective reasoning} in which the model autonomously drives a sample-adaptive, progressively coherent analysis.
This calls for \emph{self-elicited multi-perspective reasoning} in which the model autonomously explores and integrates different analytical perspectives on its own and progressively builds a sample-adaptive, coherent analysis.
In this work, we make the first attempt to equip multimodal sarcasm detection with \emph{self-elicited multi-perspective reasoning}, 
% where a single model dynamically generates the analytical perspectives needed for each sample and progressively integrates them into a coherent analysis, 
% % with each perspective building on the evidence and gaps revealed by preceding analyses.
% % unresolved questions，remaining ambiguities，unaddressed aspects
% with each perspective building on the findings established by preceding analyses and the unaddressed aspects they reveal.
a capability that enables the model to dynamically generate the analytical perspectives required for each input sample and progressively integrate them into a coherent analysis, where each perspective builds upon evidence uncovered by preceding analyses.
% Achieving this goal poses two practical challenges.
Nevertheless, achieving this goal raises two considerations.
% First, a single model reasoning on its own 
% First, even a model capable of spontaneous multi-perspective reasoning may overlook critical perspectives or produce flawed analyses, and growing evidence shows that LLMs cannot reliably self-correct without external signals~\cite{huang2024large,kamoi2024when,tsui2025selfcorrection,zhan2026selfcorrect,li2026decomposing}.
% An independent mechanism is therefore needed to assess and guide the reasoning.
% % Second, this reasoning capability cannot be elicited through prompting alone; it must be explicitly trained into the model, yet existing sarcasm datasets provide only binary labels with no reasoning annotations.
% Second, sarcasm involves subtle, entangled cues that demand fine-grained semantic sensitivity beyond what prompting alone can reliably provide; task-specific training is needed to internalize this capability, yet existing sarcasm datasets offer only binary labels with no reasoning annotations.
% Sarcasm relies on subtle cues that are often implicit and context-dependent, yet existing sarcasm datasets provide only binary labels without reasoning annotations, leaving no supervision signal for training.
\textbf{\ding{182} Absence of process-level supervision.} 
Internalizing the self-elicited multi-perspective reasoning capability in a single model requires training with explicit supervision over the reasoning process, yet existing sarcasm datasets~\cite{qin2023mmsd2,desai2022nice}
% provide only binary labels without reasoning annotations, leaving no supervision signal for training.
provide only binary labels or brief explanation of the intended meaning, with no annotations capturing the underlying analytical process.
% \textbf{① Reasoning quality is hard to guarantee.} 
% \textbf{\ding{183} Unreliable reasoning without correction.} Spontaneous multi-perspective reasoning is inherently prone to flaws: a single model may overlook critical perspectives, produce incoherent chains, or produce flawed analyses. Moreover, self-correction is unreliable without external signals~\cite{huang2024large,kamoi2024when,tsui2025selfcorrection,zhan2026selfcorrect,li2026decomposing}. 
\textbf{\ding{183} Reasoning reliability under subtle cues.}
Because sarcastic intent often depends on subtle, implicit, and cross-modal cues, a model equipped with self-elicited multi-perspective reasoning capability may still leave important aspects unaddressed, misinterpret weak signals, or fail to resolve the underlying incongruity, making the resulting reasoning incomplete or flawed.

In view of these, we present \textbf{ProCrit}, a \textbf{Pro}posal--\textbf{Crit}ic two-agent framework comprising a \emph{proposal agent} that performs multi-perspective reasoning and a \emph{critic agent} that provides external evaluation and targeted revision guidance, building on three components.
% \textbf{\ding{182} Dynamic-role agentic rollout}.
% To provide process-level training supervision, we design a \emph{dynamic-role agentic rollout} that synthesizes process-level reasoning annotations by having a strong teacher model sequentially spawn analytical roles, each contributing a distinct perspective within a shared context. Each role has full visibility of all preceding analyses and autonomously determines what perspective is still needed and when analysis is sufficient. 
\textbf{\ding{182} Process-level reasoning annotation synthesis}.
To provide process-level training supervision, we synthesize reasoning annotations through a \emph{dynamic-role agentic rollout}, where a strong vision-language model sequentially spawns analytical roles, each contributing a distinct perspective within a shared context. Each role has full visibility of all preceding analyses and autonomously determines what perspective is still needed and when analysis is sufficient. 
% The resulting analyses are then flattened to distill the sequential multi-role collaborative dynamics into a single student model, effectively transforming explicit cross-role context propagation into implicit information flow through hidden states.
The resulting multi-role collaborative dynamics is then flattened into a unified reasoning sequence and distilled into a student model, effectively transforming explicit cross-role context propagation into implicit information flow, eliminating the need to re-encode prior context and yielding rich process-level supervision.
\textbf{\ding{183} Draft--critique--revise paradigm}.
To improve reasoning reliability, we introduce a \emph{draft--critique--revise} paradigm
that decouples generation from evaluation rather than relying on self-correction, which prior work shows can be unreliable without external signals~\cite{huang2024large,kamoi2024when,tsui2025selfcorrection,zhan2026selfcorrect,li2026decomposing,arimbur2026many}: the proposal agent first drafts an initial multi-perspective reasoning; the independent critic agent then evaluates it and provides targeted natural-language feedback that points out specific deficiencies; the proposal agent revises its reasoning from scratch under this feedback, enabling directed improvement rather than uninformed retry.
\textbf{\ding{184} Mutual-refinement training strategy}.
We further develop a \emph{mutual-refinement} training strategy  
% that co-optimizes both agents: 
where the two agents are alternately optimized through reciprocal training signals: 
the proposal agent's drafting and revision abilities are jointly optimized via dual-stage reinforcement learning using the critic agent's feedback as training signal, while the proposal's revision outcomes reciprocally refine the critic---grounding its feedback quality in actual revision effectiveness.

% \emph{mutual-refinement} training framework where the two agents are alternately optimized through reciprocal training signals

% This closes the optimization loop between the two agents: each agent is refined using the other's outcomes as training signal, while the alternating freeze optimize pattern avoids the co-adaptation instabilities of simultaneous training.

% Experiments on MMSD2.0 and MMSD demonstrate that RECAST achieves state-of-the-art performance among reasoning-based approaches, substantially outperforming prompting baselines built on models with up to twice the parameters.

Our contributions are summarized as follows:
\begin{itemize}
    \item We introduce \textbf{ProCrit}, a two-agent framework that reformulates multimodal sarcasm detection from manually prescribing perspectives to self-elicited multi-perspective reasoning, where analytical perspectives are generated adaptively and integrated progressively.
    % each new perspective is generated based on the evidence and gaps revealed by preceding analyses, and thereby builds progressively on earlier findings
    % \item We design a \emph{draft--critique--revise} paradigm where an independent critic agent evaluates the initial reasoning and provides targeted feedback, enabling directed revision rather than uninformed retry.
    % \item We propose a \emph{dynamic-role agentic rollout} for reasoning data synthesis that autonomously generates sample-adaptive, progressively coherent multi-perspective reasoning without predefined role taxonomies or routing rules.
% \item We introduce a dynamic-role agentic rollout to generate process-level supervision for sample-adaptive reasoning without predefined role taxonomies or routing rules.
    % \item We introduce a \emph{dynamic-role agentic rollout} that synthesizes process-level reasoning annotations by generating progressive, sample-specific analytical trajectories.
        \item We introduce a process-level reasoning annotation synthesis strategy based on \emph{dynamic-role agentic rollout}, which generates progressive, sample-specific analytical trajectories.
    % \item We develop a \emph{mutual-refinement} training framework that jointly optimizes reasoning and revision via dual-stage reinforcement learning using the critic's feedback, with the critic reciprocally refined through actual revision outcomes.

    % \item We propose a critic-guided draft--critique--revise framework with mutual-refinement training to jointly improve reasoning, revision, and critique quality.
 \item We propose a critic-guided \emph{draft--critique--revise} paradigm with \emph{mutual-refinement training strategy}, which improves reasoning reliability through external critique while optimizing revision and critique quality in a reciprocal optimization process.

\end{itemize}

\section{Method}
\label{sec:method}

We develop \textbf{ProCrit}, a Proposal-Critic two-agent framework for multi-perspective reasoning in multimodal sarcasm detection, comprising a proposal agent that drafts and revises multi-perspective reasoning and a critic agent that provides external evaluation and targeted revision guidance.
The framework contains three components:
% (1) a \emph{dynamic-role agentic rollout} synthesizes process-level reasoning annotations by dynamically generating analytical roles based on the findings established by preceding analyses and the remaining unaddressed aspects.
(1) \emph{process-level reasoning annotation synthesis} uses a dynamic-role agentic rollout to dynamically generate analytical roles based on the findings established by preceding analyses and the remaining unaddressed aspects;
% , without relying on predefined role taxonomies or routing rules.
% each new perspective is generated based on the evidence and gaps revealed by preceding analyses, and thereby builds progressively on earlier findings
(2) a \emph{draft--critique--revise paradigm}  improves reasoning reliability by providing external evaluation and targeted guidance for revision;
% (3) a \emph{mutual-refinement} training framework jointly improves proposal drafting, feedback-guided revision, and critic feedback quality.
(3) a \emph{mutual-refinement training strategy} that alternately optimizes the two agents through reciprocal training signals: the critic agent's feedback improves the proposal agent's drafting and revision capabilities, while the proposal agent's revision outcomes reciprocally refine the critic agent.
% Figure~\ref{fig:framework} illustrates the overall architecture.

% 生成不同的角色 dynamic-role agentic rollout
% **Greater adaptability to sample-specific sarcasm mechanisms** — 每个样本的 step title 序列不同，因为讽刺机制不同需要的分析角度不同
% 1. **Reduced dependence on manually designed agent taxonomies and routing rules** — 不用人为定义"哪些专家角色"和"什么时候调用谁"
% 2. **Context-aware progressive planning** — 后续步骤基于前面已有的分析发现来选择新角度（prompt 要求 "add a different angle"），不是独立专家各做各的，没有cross-perspective interaction”
 % data framework strategy
 \begin{figure}
    \centering
    \includegraphics[width=1\linewidth]{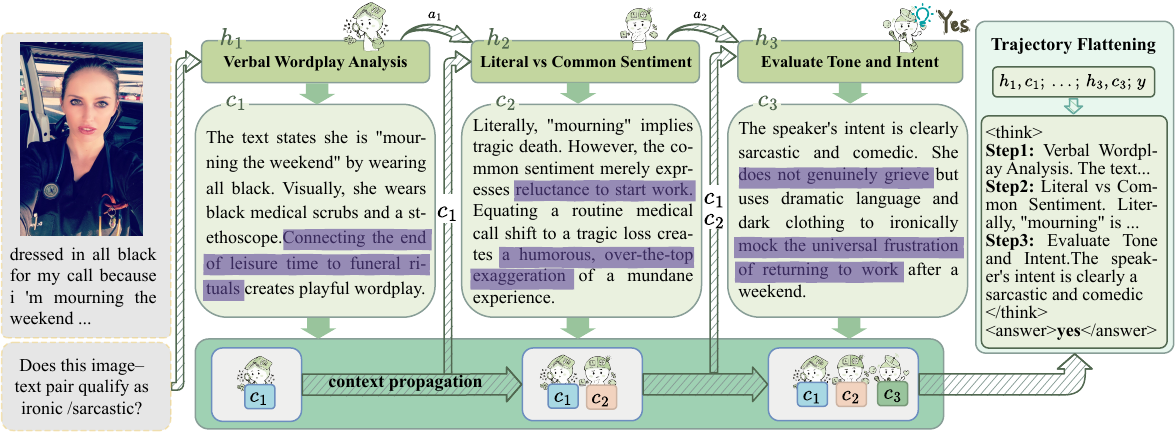}
    \caption{Illustration of the reasoning annotation synthesis process.}
    \label{fig:Synthesis}
    \vspace{-11pt}
\end{figure}
% ------------------------------------------------------------------
%  3.1 — Dynamic-Role Agentic Rollout
% ------------------------------------------------------------------
\subsection{Process-Level Reasoning Annotation Synthesis}
\label{sec:data_synthesis}

% Existing sarcasm corpora provide only binary labels without reasoning 
% annotations, and existing multi-role prompting approaches assign fixed 
% analytical roles whose outputs are aggregated independently. To obtain 
% training data with coherent, sample-adaptive multi-perspective 
% reasoning, we design a dynamic-role agentic rollout in which a strong 
% vision-language model~$\mathcal{T}$ sequentially spawns analytical 
% roles within a shared context.

% Existing sarcasm corpora provide only binary labels, with no annotations capturing the underlying analytical process.
While existing sarcasm corpora~\cite{qin2023mmsd2,desai2022nice} offer binary labels or brief explanations of the intended meaning, they lack process-level annotations that capture the reasoning steps or critical cues through which sarcastic intent is identified, as illustrated by the comparison in Figure~\ref{fig:comparison}.
% , while existing multi-role prompting approaches assign fixed analytical roles whose outputs are aggregated independently. 
To obtain process-level supervision for coherent, sample-adaptive multi-perspective reasoning, we design a \emph{dynamic-role agentic rollout} in which strong vision-language model, denoted as~$\mathcal{T}$, sequentially spawns analytical roles within a shared context.

\textbf{Dynamic-role agentic rollout.}
Given an image-text pair $(I, Q)$, the model~$\mathcal{T}$ generates a reasoning trajectory by iteratively producing analytical roles, each contributing a distinct perspective. 
The $t$-th role yields a
structured output $\mathbf{r}_t=(h_t, c_t, a_t)$, where
$h_t$ is its analytical perspective, $c_t$ is evidence-focused
analysis, and $a_t$ is a sufficiency judgment indicating whether the accumulated analyses suffice for a final prediction.
Each role is conditioned on the full history of preceding roles:
\begin{equation}
(h_t, c_t, a_t) = \mathcal{T}\big(I,\, Q,\,
[(h_1, c_1), \ldots, (h_{t-1}, c_{t-1})],\, a_{t-1}\big).
\label{eq:rollout}
\end{equation}
% Because each successive role has full visibility of all prior
% analyses, it can identify gaps in earlier reasoning, avoid redundant
% analysis, and adaptively determine both the direction and depth of
% further investigation. 
% Each successive role is instructed to introduce a different angle of analysis, enabling the model to progressively build a multi-faceted understanding.
% With full visibility of prior analyses, each successive role can build on established findings, identify unaddressed aspects, avoid redundancy, and adaptively guide further investigation, while contributing a distinct perspective that collectively builds a multi-faceted understanding.
% This design makes both the number and nature of roles fully adaptive while eliminating the need for manually designed role taxonomies or routing rules.
% With full visibility of prior analyses, each successive role builds on established findings, identifies unaddressed aspects, and avoids redundancy.By introducing a distinct perspective, it extends the reasoning trajectory and progressively contributes to a coherent multi-perspective reasoning.
With full visibility of prior analyses, each successive role 
builds on established findings and identifies unaddressed aspects, 
progressively extending the reasoning trajectory by introducing a distinct perspective.
When $a_t$ indicates that the analyses are sufficient, the rollout
terminates and $\mathcal{T}$ produces the final prediction $\hat{y}$,
forming a trajectory $\tau=(\mathbf{r}_1,\ldots,\mathbf{r}_T,\hat{y})$.
As a result, both the number of roles $T$ and the perspectives $h_t$ $(t=1,...,T)$ they cover are fully adaptive, governed by the sufficiency judgment $a_t$ rather than manually designed role taxonomies or routing rules.

% We sample $N$ trajectories per query with stochastic decoding and retain those with correct final answers.

% 跨角色依赖以及推理逻辑已经包含在组成的推理链中
% - 上下文（image + question + 指令）只编码一次
% - 隐状态自然从 Step 1 流到 Step 2 再到 Step 3，不需要通过显式的 text token 来传递上下文——前面步骤的信息已经在 KV cache / hidden states 中了  
% - 所有推理步骤在一次 autoregressive generation 中连续输出

% 显式的跨角色协作被内化为经由隐状态传递的隐式信息流，使得在单次自回归前向传播过程中即可实现多视角推理——在此过程中，共享上下文仅被编码一次。

\textbf{Trajectory flattening.}
The agentic rollout above produces a multi-turn trajectory $\tau$ in which each role conditions on the full history of all preceding roles.
However, directly using this multi-turn format for training would require the shared multimodal context to be re-encoded at every turn, leading to redundant computation and quadratically growing input length.
We therefore flatten $\tau$ into a single sequence by concatenating all role outputs in order, followed by the final prediction:
% The trajectory $\tau$ records the ordered role outputs together with
% the final prediction. We then flatten $\tau$ into a unified
% reasoning sequence that encodes the full multi-role analytical
% progression:
\begin{equation}
  \mathbf{s} = [\mathbf{r}_1;\, \mathbf{r}_2;\, \ldots;\, 
  \mathbf{r}_T;\, \hat{y}]
  = [h_1, c_1;\, h_2, c_2;\, \ldots;\, h_T, c_T;\, \hat{y}],
  \label{eq:flatten}
\end{equation}
where $[\,\cdot\,;\,\cdot\,]$ denotes sequence concatenation.
% The resulting sequence preserves the progressive cross-perspective dependencies in a form directly learnable by standard autoregressive training.
% Essentially, explicit cross-role context propagation is effectively transformed into implicit state propagation, thereby avoiding repeated encoding of the shared context and removing the need to re-encode prior role outputs as explicit history at each step.
The resulting data provide rich process-level supervision that preserves multi-role analytical dependencies in a single reasoning sequence: a model trained on these sequences learns to produce all analytical perspectives in one autoregressive pass, 
% where information from earlier roles flows to later ones through hidden states, eliminating repeated context re-encoding.
where information from earlier roles flows to later ones through hidden states within the same forward pass, eliminating the need to re-encode prior outputs at each turn.

% avoiding repeated role-level interaction.
 % within a single continuous generation
% In effect, explicit cross-role context propagation in the teacher rollout
% is distilled into implicit state propagation within the proposal agent,
% enabling multi-perspective reasoning to be executed as a continuous
% single-pass generation rather than repeated role-level interaction.
% In effect, explicit cross-role context propagation is transformed into implicit state propagation, avoiding repeated context encoding while allowing earlier analyses to condition later perspectives within a single continuous generation.
% Within this continuous sequence, the shared multimodal context is established once, while earlier analyses are carried forward through the evolving prefix and model state rather than re-encoded as an explicit cross-role history at each step.

% \begin{figure}[h]
%     \centering
%     \includegraphics[width=1\textwidth]{framework.pdf}
%     \caption{Overview of ProCrit. 
% \textbf{Top:} Reasoning annotation synthesis via dynamic-role 
% agentic rollout and trajectory flattening. 
% \textbf{Bottom:} Mutual-refinement training of the critic agent 
% (left) and the proposal agent (right).}
%     \label{fig:framework}
% \end{figure}

% ！！！不是堆实现细节，而是每个设计都按“为什么需要这个设计 → 它怎么抽象地工作 → 带来什么建模收益”来写。
% --------

%==================================================================
%  3.2 — Draft--Critique--Revise Paradigm
% ==================================================================
\vspace{-4pt}

\subsection{Draft--Critique--Revise Paradigm}
\label{sec:framework}
\vspace{-4pt}

% However, a single-pass draft may still contain reasoning errors that the model itself cannot reliably detect~\cite{huang2024large}.
% To enable directed revision, we pair the proposal agent with an independent critic agent, forming a \emph{draft--critique--revise} cycle in which the critic identifies specific deficiencies in the draft and provides natural-language guidance for revision.

% To further improve reasoning reliability, we introduce an independent critic agent that assesses draft reasoning and provides targeted revision guidance, forming a \emph{draft--critique--revise} cycle between two collaborating agents.

With the process-level supervision above, the proposal agent is trained to generate sample-adaptive multi-perspective reasoning for a given sample (details in \S\ref{sec:training}).
However, since sarcastic intent often hinges on subtle and implicit cues, such single-pass reasoning may still overlook critical perspectives or misinterpret weak signals, and growing evidence shows that models cannot reliably detect flaws in their own reasoning~\cite{huang2024large}.
Therefore, to further improve reasoning reliability, we introduce an independent critic agent to decouple reasoning generation from reasoning evaluation, forming a \emph{draft--critique--revise} paradigm in which the critic agent identifies specific reasoning deficiencies and provides targeted natural-language feedback for directed revision, as illustrated in figure~\ref{fig:paradigm}.
% Together, the proposal and critic agents form a \emph{draft–critique–revise} paradigm: the proposal agent first drafts an initial reasoning; the critic agent then identifies specific deficiencies and provides targeted natural-language feedback; finally, the proposal agent performs a directed revision based on this critique, as illustrated in figure~\ref{fig:paradigm}.

% To further improve reasoning reliability, we therefore introduce an independent critic agent to decouple reasoning generation from reasoning assessment, forming a \emph{draft--critique--revise} cycle in which the proposal agent first produces an initial draft reasoning, the critic identifies its deficiencies, and the proposal agent then revises its reasoning under explicit guidance, as illustrated in Figure~\ref{fig:paradigm}.

\begin{wrapfigure}{r}{0.52\textwidth}  % r 表示右侧，0.4 表示占页面宽度的 40%
  \vspace{-18pt}
  \centering
  \includegraphics[width=0.52\textwidth]{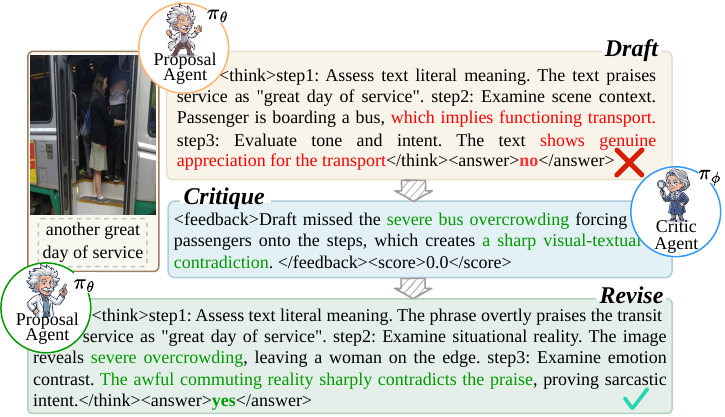} 
  % \vspace{-5pt}
  \caption{Illustration of the draft--critique--revise paradigm.}
  \label{fig:paradigm}
  \vspace{-13pt}
\end{wrapfigure}

\textbf{Draft.}
\label{sec:proposal_agent}
Given an image-text pair $(I, Q)$, the proposal agent~$\pi_\theta$ operates in \emph{draft} mode and generates an initial multi-perspective reasoning~${\bar{d}}$ together with a binary prediction~$\hat{y}_{\mathrm{draft}}$ under a drafting prompt~$p_{\mathrm{draft}}$:
\begin{equation}
  ({\bar{d}},\, \hat{y}_{\mathrm{draft}}) = \pi_\theta(I,\, Q;\, p_{\mathrm{draft}}).
  \label{eq:draft}
\end{equation}
The draft reasoning~${\bar{d}}$ follows the flattened dynamic-role format learned from the synthesized trajectories: a coherent sequence of sample-specific analytical perspectives within \texttt{<think>}$\cdots$\texttt{</think>} tags, followed by the predicted sarcasm label in \texttt{<answer>}$\cdots$\texttt{</answer>} tags.

% Thus, the draft stage provides both an interpretable reasoning object to be inspected and an initial decision that can later be corrected.

% ------------------------------------------------------------------
\textbf{Critique.}
\label{sec:critic_agent}
% ProCrit subjects the proposal draft to an independent external assessment before revision.
% The critic agent~$\pi_\phi$, a separately trained model, evaluates the draft reasoning and produces two complementary outputs:
% The critic agent~$\pi_\phi$ is trained separately from the proposal agent and receives the original image-text pair $(I, Q)$ together with the proposal's draft~$\mathbf{d}$:
% To further improve the reliability of the initial draft reasoning~$\mathbf{d}$, an independent critic agent~$\pi_\phi$ is introduced to provide external assessment: 
To further refine the initial draft reasoning~${\bar{d}}$, the critic agent~$\pi_\phi$ evaluates it against the original multimodal evidence under a critic prompt~$p_{\mathrm{critic}}$:
% the draft reasoning~$\mathbf{d}$ is then evaluated by an independent critic agent~$\pi_\phi$, which takes the original image-text pair $(I, Q)$ together with the draft reasoning~$\mathbf{d}$ as input and produces two complementary signals: 
\begin{equation}
  (f,\, s) = \pi_\phi(I,\, Q,\, {\bar{d}};\, p_{\mathrm{critic}}),
  \label{eq:critique}
\end{equation}
where $f$~is a natural-language feedback within \texttt{<feedback>}$\cdots$\texttt{</feedback>} tags that identifies specific reasoning deficiencies and provide targeted revision guidance,
and $s \in \{0, 1, 2\}$~is a quality score within \texttt{<score>}$\cdots$\texttt{<score>} tags, providing an estimate of overall reasoning quality.
% The feedback $f$ conditions the subsequent revision (Eq.~\ref{eq:revise}), while the score serves as a reward signal for proposal agent optimization (\S\ref{sec:proposal_rl}).

% ------------------------------------------------------------------
\textbf{Revise.}
% The same proposal agent~$\pi_\theta$ then switches to \emph{revise} mode.
% Conditioned on the original input, its previous draft, and the critic feedback, it generates a revised reasoning from scratch:
Guided by the feedback~$f$, the same proposal agent~$\pi_\theta$ then switches to \emph{revise} mode and generates a revised reasoning from scratch under a revision prompt~$p_{\mathrm{revise}}$ (see Appendix~\ref{app:prompts}):
\begin{equation}
  % (\mathbf{d}',\, \hat{y}_{\mathrm{r}}) = \pi_\theta(I,\, Q,\, \mathbf{d},\, f).
  ({d},\, \hat{y}_{\mathrm{revise}}) = \pi_\theta(I,\, Q,\, {\bar{d}},\, f;\, p_{\mathrm{revise}}).
  \label{eq:revise}
\end{equation}
The revised reasoning~$\mathbf{d}$ follows the same structured format as the draft reasoning~$\mathbf{\bar{d}}$, and the revised output $({d}, \hat{y}_{\mathrm{revise}})$ is taken as the final result of the draft--critique--revise pipeline, while the draft output $({\bar{d}}, \hat{y}_{{draft}})$ remains available as the first-pass result for training and analysis.
\subsection{Mutual-Refinement Training}
\label{sec:training}

% The draft--critique--revise paradigm requires both agents to perform well: the proposal agent must produce coherent reasoning and effectively incorporate feedback, while the critic must provide feedback that is genuinely actionable rather than superficially plausible.
% We address this through a \emph{mutual-refinement} training framework in which the two agents serve as reciprocal training signals for each other.
% Specifically, the critic's feedback drives the proposal agent's revision training via dual-stage reinforcement learning (\S\ref{sec:proposal_rl}), while the proposal agent's revision outcomes in turn refine the critic---grounding its feedback quality in actual revision effectiveness (\S\ref{sec:critic_rl}).
% This creates a closed optimization loop: critique quality is directly measured by whether it leads to improved reasoning, ensuring that the guidance is genuinely useful.

% We further develop a \emph{mutual-refinement} training strategy that alternately optimizes the two agents through reciprocal training signals.

We further develop a \emph{mutual-refinement} training strategy that alternately optimizes the two agents, each serving as a training signal for the other.
Specifically, when optimizing the proposal agent, the critic agent is frozen and provides quality scores and targeted feedback as reward signals for dual-stage reinforcement learning for drafting and revision (\S\ref{sec:proposal_rl});
when optimizing the critic agent, the proposal agent is frozen and its revision outcomes measure whether the critic's feedback genuinely improves reasoning (\S\ref{sec:critic_rl}).
% This alternating freeze--optimize design refines both agents while avoiding the co-adaptation instabilities of simultaneous training.
% This alternating freeze--optimize design enables mutual refinement while avoiding the co-adaptation instabilities of simultaneous training.
% The following subsections detail each stage in turn.
Through this alternation, both agents are progressively refined: the proposal agent produces stronger reasoning, and the critic agent provides more actionable feedback.
% ------------------------------------------------------------------
\subsubsection{Initialization}
\label{sec:sft_init}

% Both agents are initialized via supervised fine-tuning (SFT) to establish basic capabilities for sample-adaptive multi-perspective reasoning and the critic for quality scoring and targeted feedback generation.
% The training data for the two agents are constructed through an interleaved process: the proposal agent's draft data come from the dynamic-role synthesis pipeline (\S\ref{sec:data_synthesis}); these drafts are then scored by the teacher model~$\mathcal{T}$ to produce critic training data; and the resulting feedback is in turn used to elicit revision examples for the proposal agent.
% % This interleaved construction ensures that both agents are initialized on data that reflects the draft--critique--revise interaction pattern they will encounter during RL.
% Implementation details are provided in \S\ref{sec:setup}.

Both agents are initialized via Supervised Fine-Tuning (SFT) to establish basic capabilities, with training data constructed through an interleaved process. 
The proposal agent is first trained on the synthesized reasoning trajectories (\S\ref{sec:data_synthesis}) for sample-adaptive multi-perspective reasoning; 
the trained proposal agent then generates drafts on the training set, which are evaluated by the model~$\mathcal{T}$ to produce training examples for the critic agent; 
the resulting feedback from the trained critic agent is in turn used to prompt~$\mathcal{T}$ for revision, which serve as the proposal agent's SFT data for the revise mode.
% revise-mode SFT data for the proposal agent 
% the resulting critic feedback is in turn paired with the drafts to construct revision examples for the proposal agent. 
Implementation details are provided in \S\ref{sec:setup}.

% ------------------------------------------------------------------
\subsubsection{Proposal Agent: Dual-stage Reinforcement Learning}
\label{sec:proposal_rl}

The proposal agent is optimized through a dual-stage Group Relative Policy Optimization (GRPO) objective that jointly trains its drafting and revising capabilities.
During training, the critic agent is frozen and deployed as an inference service.
% , supplying both quality scores and natural-language feedback for each draft.

\textbf{Stage~1: Draft optimization.}
For each image-text pair~$(I, Q)$, the proposal agent generates $G$ draft completions $\{d_1, \ldots, d_G\}$.
The critic agent evaluates each draft $d_i$, producing the corresponding quality score~$s_i$ and natural-language feedback~$f_i$.
% The draft reward combines three reward components:
Each draft is assigned a composite reward:
\begin{equation}
  r_{\mathrm{draft}}(d_i) = r_{\mathrm{acc}}(d_i) + r_{\mathrm{fmt}}(d_i) +  r_{\mathrm{eval}}(d_i),
  \label{eq:draft_reward}
\end{equation}
% where $r_{\mathrm{acc}} \in \{0, 1\}$ is a binary accuracy reward, $r_{\mathrm{fmt}} \in \{0, 1\}$  is a format reward, 
where $r_{\mathrm{acc}} \in \{0, 1\}$ is a binary accuracy reward indicating whether the predicted label matches the ground truth, $r_{\mathrm{fmt}} \in \{0, 1\}$ is a format reward verifying adherence to the required output format,
$r_{\mathrm{eval}}(d_i)=\tilde{s}_i
$ is a critic evaluation reward given by the normalized quality score.

\textbf{Stage~2: Feedback-guided revision optimization.}
% From the $G$ drafts, $K$ parent drafts are selected via correctness-balanced sampling.
% From the $G$ drafts, we select $K$ parent drafts by random sampling with equal probability over correct and incorrect draft pools.
% From the $G$ drafts, we select $K$ parent drafts by randomly sampling equally from the correct and incorrect draft pools.
% From the $G$ drafts, $K$ parents are selected via balanced sampling: each parent is drawn with equal probability from the correct and incorrect draft pools.
From the $G$ drafts, $K$ parents are selected via balanced sampling to include both correct and incorrect drafts.
For each selected parent~$d_k$ with critic feedback~$f_k$, the proposal agent generates $M$ revised completions $\{o_{k,1}, \ldots, o_{k,M}\}$ by rewriting the draft reasoning from scratch under the feedback.
% From the $G$ drafts, we select $K$ parent drafts by first choosing whether to sample from the correct or incorrect drafts with equal probability, and then randomly selecting a draft from the chosen group.
% From the $G$ drafts, we select $K$ parent drafts using correctness-balanced sampling: when both correct and incorrect drafts are available, each parent is drawn from either group with equal probability; otherwise, sampling falls back to the available group.
The revision reward is defined as:
\begin{equation}
  r_{\mathrm{revise}}(o_{k,j}) = r_{\mathrm{acc}}(o_{k,j}) + r_{\mathrm{fmt}}(o_{k,j}) + r_{\mathrm{imp}}(o_{k,j},\, d_k),
  \label{eq:react_reward}
\end{equation}
where the improvement reward $r_{\mathrm{imp}}$ is designed to capture the marginal effect of revision relative to the draft and $y^*$ denotes the gold label:
% where the improvement reward is defined as:
\begin{equation}
\begin{aligned}
r_{\mathrm{imp}}(o, d) =
  \begin{cases}
    +1 & \text{if } \hat{y}_{\mathrm{draft}} \neq y^* \text{ and } \hat{y}_{\mathrm{revise}} = y^* \quad\text{(fix)}\\
    -1 & \text{if } \hat{y}_{\mathrm{revise}} \neq \hat{y}_{\mathrm{draft}} \text{ and } \hat{y}_{\mathrm{revise}} \neq y^* \quad\text{(damage)} \\
    \phantom{+}0  & \text{otherwise}
  \end{cases}
\end{aligned}
\label{eq:improve_reward}
\end{equation}
% We design $r_{\mathrm{imp}}$ to capture the marginal effect of revision relative to the draft: 
Since correctness is already captured by $r_{\mathrm{acc}}$, $r_{\mathrm{imp}}$ gives additional credit only to genuine corrections, penalizes harmful changes, and remains neutral for rewrites without a correctness gain.

\begin{figure}
    \centering
    \includegraphics[width=1\linewidth]{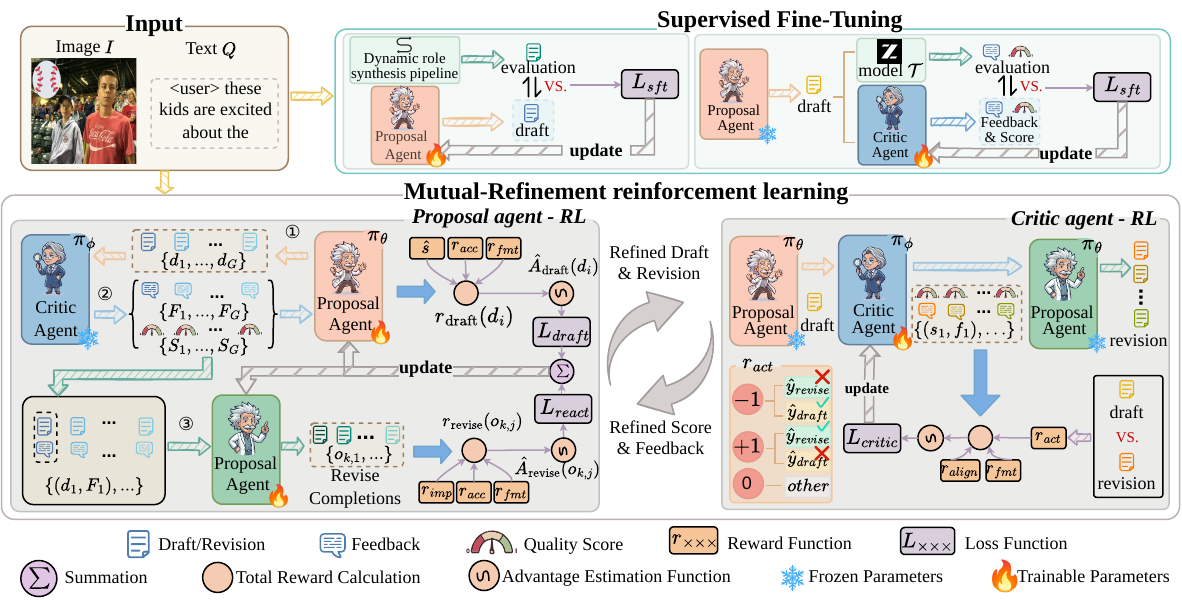}
    % \caption{Overview of the mutual-refinement training pipeline. \textbf{Top:} Both agents are initialized via supervised fine-tuning---the proposal agent on synthesized reasoning annotations and the critic agent on quality assessments of proposal drafts. \textbf{Bottom left:} Critic RL optimizes the critic agent while the proposal agent is frozen; the proposal's revision outcomes serve as outcome-grounded signals to reward actionable feedback. \textbf{Bottom right:} Proposal RL optimizes the proposal agent in a dual-stage process (draft and revision) while the critic agent is frozen, providing quality scores and targeted feedback.}

    % \caption{Overview of the mutual-refinement training pipeline. \textbf{Top:} Both agents are initialized via supervised fine-tuning. \textbf{Bottom left:} The critic agent is optimized while the proposal is frozen, using revision outcomes as reward signals. \textbf{Bottom right:} The proposal is optimized in dual stages (draft and revision) while the critic agent is frozen.}

\caption{Overview of mutual-refinement training. \textbf{Top:} 
Both agents are initialized via SFT. \textbf{Bottom left:} 
The critic is optimized while the proposal is frozen, using 
revision outcomes as rewards. \textbf{Bottom right:} The 
proposal is optimized in dual stages (draft and revision) 
while the critic is frozen.}

    \label{fig:Mutual_Refinement}
    \vspace{-8pt}
\end{figure}

\textbf{Dual-stage GRPO.}
% Dual-stage GRPO advantage estimation and objective
% Rather than treating drafting and revision as two independent reinforcement learning tasks, we optimize them as a coupled two-stage rollout.
% The two stages share a single training rollout and a joint gradient update.
Rather than optimizing drafting and revision separately, we couple them into a single training rollout that produces a joint gradient update.
% In the draft stage, $G$ candidate completions for the same input are compared against each other; in the revision stage, $M$ feedback-guided revisions of the same parent draft are compared within each parent group.
Let $\mathcal{D}=\{d_i\}_{i=1}^{G}$ denote the draft candidate set and $\mathcal{O}_k=\{o_{k,j}\}_{j=1}^{M}$ denote the revision candidate set for parent draft~$d_k$.
We compute group-relative advantages within each set:
\begin{equation}
\begin{aligned}
\hat{A}_{\mathrm{draft}}(d_i)
&= \frac{r_{\mathrm{draft}}(d_i)-\mu}{\sigma+\epsilon},
& \mu &= \operatorname{mean}_{d\in\mathcal{D}} r_{\mathrm{draft}}(d),
& \sigma &= \operatorname{std}_{d\in\mathcal{D}} r_{\mathrm{draft}}(d),\\
\hat{A}_{\mathrm{revise}}(o_{k,j})
&= \frac{r_{\mathrm{revise}}(o_{k,j})-\mu_k}{\sigma_k+\epsilon},
& \mu_k &= \operatorname{mean}_{o\in\mathcal{O}_k} r_{\mathrm{revise}}(o),
& \sigma_k &= \operatorname{std}_{o\in\mathcal{O}_k} r_{\mathrm{revise}}(o).
\end{aligned}
\label{eq:advantage}
\end{equation}
% Here, $\epsilon$ is a small numerical constant, set to $10^{-4}$ in implementation.
Let $\Phi(\mathcal{C}, \hat{A})$ denote the GRPO clipped policy-ratio objective evaluated over a candidate set~$\mathcal{C}$ with its group-relative advantages~$\hat{A}$.
The proposal agent is trained with a single dual-stage objective:
\begin{equation}
\mathcal{L}_{\mathrm{proposal}}
= (1-\lambda)\Phi(\mathcal{D}, \hat{A}_{\mathrm{draft}})
+ \lambda \frac{1}{K}\sum_{k=1}^{K}\Phi(\mathcal{O}_k, \hat{A}_{\mathrm{revise}}),
\label{eq-10}
\end{equation}
where $\lambda \in [0,1]$ controls the relative emphasis on feedback-guided revision.
% Because both stages update the same parameters, revision training simultaneously reinforces drafting quality---the model learns not only to revise effectively but also to produce better initial reasoning that requires less correction.
% Because both stages update the same parameters, feedback-guided revision also provides supervision for draft generation: by learning how flawed drafts are corrected, the model also learns to avoid similar deficiencies in its initial reasoning.
% Since both stages update the same parameters, the revision gradient also shapes draft-stage behavior: by learning to correct flawed reasoning under feedback, the model simultaneously learns to avoid such flaws in its initial drafts.
Since both stages share parameters, revision training 
simultaneously strengthens drafting: learning to correct flaws 
under feedback helps the model avoid similar flaws in its 
initial drafts.

\subsubsection{Critic agent: Reinforcement Learning}
\label{sec:critic_rl}

After supervised fine-tuning, the critic agent is further optimized via GRPO to improve its scoring calibration and feedback utility.
% scoring accuracy
% scoring calibration
% the alignment between its scores and proposal correctness
For each training sample containing an image-text pair~$(I, Q)$ and a draft reasoning, $G$ completions are sampled and rewarded by the composite reward $r_{\mathrm{critic}} = r_{\mathrm{fmt}} + r_{\mathrm{align}} + r_{\mathrm{act}}$. The three components are defined as follows.

\textbf{Format reward~$r_{\mathrm{fmt}}$.}
A binary reward verifying that the output contains valid \texttt{<feedback>}$\cdots$\texttt{</feedback>} and \texttt{<score>}$\cdots$\texttt{</score>} tags with non-empty content.
% A binary reward equal to 1 if the completion contains valid \texttt{<feedback>}$\cdots$\texttt{</feedback>} and \texttt{<score>}$\cdots$\texttt{</score>} tags with non-empty feedback and a score in the valid range, and 0 otherwise.

\textbf{Score alignment reward~$r_{\mathrm{align}}$.}
Let $s$ denote the critic's predicted score, $\hat{y}$ the proposal's answer:
\begin{equation}
  r_{\mathrm{align}} =
  \begin{cases}
    s - 1  & \text{if } \hat{y} = y^* \\
    1 - s  & \text{if } \hat{y} \neq y^*.
  \end{cases}
  \label{eq:score_align}
\end{equation}
This rewards the critic for assigning high scores to correct proposals and low scores to incorrect ones.
% The gold label is used only in the reward computation, not as input to the critic.

\textbf{Feedback actionability reward~$r_{\mathrm{act}}$.}
The critic's feedback is sent to a frozen proposal agent deployed as a separate inference service, which generates a corresponding revision.
% Let $\hat{y}_{\mathrm{draft}}$ and $\hat{y}_{\mathrm{revise}}$ denote the draft and revision:
\begin{equation}
  r_{\mathrm{act}} =
  \begin{cases}
    +1 & \text{if } \hat{y}_{\mathrm{draft}} \neq y^* \text{ and } \hat{y}_{\mathrm{revise}} = y^* \quad\text{(fix)}\\
    -1 & \text{if } \hat{y}_{\mathrm{revise}} \neq \hat{y}_{\mathrm{draft}} \text{ and } \hat{y}_{\mathrm{revise}} \neq y^* \quad\text{(damage)} \\
    \phantom{+}0  & \text{otherwise}.
  \end{cases}
  \label{eq:feedback_act}
\end{equation}
We design $r_{\mathrm{act}}$ to reward feedback only when it provides
verifiable corrective value. Feedback without observable improvement is therefore treated as neutral, preventing the critic agent from learning generic or unnecessary feedback patterns.

% The total critic reward is $r_{\mathrm{critic}} = r_{\mathrm{fmt}} + r_{\mathrm{align}} + r_{\mathrm{act}}$.
% The critic agent is optimized using standard GRPO with group-wise advantage normalization over the $G$ completions per prompt.

% The total critic reward is $r_{\mathrm{critic}} = r_{\mathrm{fmt}} + r_{\mathrm{align}} + r_{\mathrm{act}}$,
% and the critic agent is optimized using standard GRPO with group-wise advantage normalization.

\vspace{-4pt}

\subsubsection{Training Pipeline}
\label{sec:training_pipeline}

% Both agents are first initialized via SFT (\S\ref{sec:sft_init}), then refined through alternating RL stages: the critic is optimized first against a frozen SFT proposal, and the proposal is subsequently optimized against the frozen GRPO-trained critic.
Both agents are first initialized via SFT (\S\ref{sec:sft_init}), then refined through alternating reinforcement learning stages.
We first optimize the critic agent with the proposal agent frozen after SFT, and then optimize the proposal agent with the critic agent frozen after its RL stage.
% This closes the optimization loop between the two agents: each agent is refined using the other's outcomes as training signal, while the alternating freeze optimize pattern avoids the co-adaptation instabilities of simultaneous training.
At each stage, one agent is frozen while the other is optimized using training signals produced by its counterpart, avoiding the co-adaptation instabilities of simultaneous training.

% % ------------------------------------------------------------------
% \subsection{Inference}
% \label{sec:inference}

% At inference time, the proposal agent processes an image-text pair in two passes:
% (1)~\textbf{Draft}: the proposal agent generates a structured reasoning chain and initial answer;
% (2)~\textbf{Critic~$\to$~React}: the frozen critic evaluates the draft and provides feedback, and the proposal agent generates a revised reasoning chain conditioned on the feedback.
% The final output is the react completion.

\vspace{-3pt}
%% ================================================================
%%  SECTION 4 — Experiments
%% ================================================================
\section{Experiments}
\label{sec:experiments}
\vspace{-4pt}

% ==================================================================
%  4.1 — Experimental Setup
% ==================================================================
\subsection{Experimental Setup}
\label{sec:setup}

\paragraph{Datasets.}
% % Following previous work
We evaluate on three widely used multimodal sarcasm detection benchmarks:
% (1)~\textbf{MMSD}~\cite{cai2019multi}, the first large-scale image-text sarcasm detection dataset collected from Twitter, containing 24317 samples with a roughly balanced label distribution;
(1)~\textbf{MMSD}~\cite{cai2019multi}, the first large-scale image-text sarcasm detection dataset collected from Twitter, containing $24,317$ samples. Its test set comprises $969$ sarcastic and $1,404$ non-sarcastic examples, respectively;
(2)~\textbf{MMSD2.0}~\cite{qin2023mmsd2}, a cleaned version of MMSD that removes the spurious cues such as hashtags and re-annotates the unreasonable samples, serving as the current standard benchmark with the same scale but more reliable labels. Its test set contains $1,037$ sarcastic and $1,372$ non-sarcastic examples, respectively;
(3)~\textbf{RedEval}~\cite{tang2024leveraging}, constructed for out-of-domain evaluation of models trained on Twitter-based datasets, and samples are sourced from Reddit, with 395 sarcastic and 609 non-sarcastic examples, respectively.

% with 395 sarcastic examples drawn from the ``sarcasm'' subreddit and 609 non-sarcastic examples from other subreddits.

% We evaluate on three multimodal sarcasm detection benchmarks:
% (1)~\textbf{MMSD}~\cite{cai2019multi}, a large-scale Twitter dataset with 24,317 samples;
% (2)~\textbf{MMSD2.0}~\cite{qin2023mmsd2}, a cleaned and re-annotated version of MMSD with more reliable labels;
% (3)~\textbf{RedEval}~\cite{tang2024leveraging}, a Reddit-sourced dataset used for out-of-domain evaluation.

% We evaluate on three benchmarks: \textbf{MMSD}~\cite{cai2019multi}, the first large-scale Twitter image-text sarcasm detection dataset with 24,635 samples; \textbf{MMSD2.0}~\cite{qin2023mmsd2}, a cleaned and re-annotated version of MMSD with more reliable labels; and \textbf{redEval}~\cite{tang2024leveraging}, which we use for out-of-domain evaluation.
% We follow the official splits for MMSD and MMSD2.0.

\textbf{Evaluation metrics.}
Following prior works~\cite{qin2023mmsd2,wang2025s3agent}, we report F1 score (\textbf{F1}), accuracy (\textbf{Acc}), precision (\textbf{P}), and recall (\textbf{R}), with F1 as the primary metric.
% , producing approximately 20,000 trajectories (step distribution: 2-step 10.9\%, 3-step 74.5\%, 4-step 14.4\%, 5-step 0.3\%).
% Training data synthesis uses a strong VLM as the teacher model, producing approximately 20,000 trajectories with a step distribution of 2-step (10.9\%), 3-step (74.5\%), 4-step (14.4\%), and 5-step (0.3\%).

\textbf{Implementation details.}
Both the proposal and critic agents are initialized from Qwen2.5-VL-7B-Instruct~\cite{bai2025qwen25vl}, with the vision encoder and multimodal projector frozen and only the language model parameters updated.
For \emph{reasoning annotations generation}, we use GLM-4.6V~\cite{glm2024chatglm} as the strong vision-language model~$\mathcal{T}$.
For \emph{supervised fine-tuning}, the proposal agent is fine-tuned on 9K generated reasoning sequences with a global batch size of 64 and a learning rate of $1 \times 10^{-5}$ for 3 epochs;
the critic agent is fine-tuned on 20K examples sourced from the proposal agent's rollouts, annotated by~$\mathcal{T}$.
For \emph{reinforcement learning}, the two agents are alternately optimized, with the frozen counterpart deployed via vLLM~\cite{kwon2023efficient} to supply training signals at each stage. The proposal agent is optimized using 2K instances with $G{=}8$, $M{=}4$, $\lambda{=}0.5$, and learning rate $3 \times 10^{-6}$ for 4 epochs. The critic agent is optimized using 5K instances with $G{=}8$ completions and learning rate $1 \times 10^{-6}$ for 4 epochs.
All training is conducted on 8 NVIDIA A800 GPUs with DeepSpeed ZeRO-3~\cite{rajbhandari2020zero} to reduce memory consumption.
More hyperparameter details are provided in Appendix~\ref{app:impl_details}.

\textbf{Evaluation Settings.}
We compare methods along two axes: training regime (\emph{zero-shot} vs.\ \emph{fine-tuned}) and reasoning perspective setting (\emph{None}, \emph{Predefined}, and \emph{Adaptive}), which correspond to the labels used in Table~\ref{tab:main_results}.
(1) \emph{None} denotes plain chain-of-thought reasoning without explicitly specified analytical perspectives; this category includes Zero-shot CoT~\cite{kojima2022large}, Automatic Prompt Engineer~\cite{zhou2023large}, Plan-and-Solve~\cite{wang2023planandsolve}, Generated Knowledge Prompting~\cite{liu2022generated}.
% , and Plain CoT Reasoner that we implement to learn to produce generic chain-of-thought analyses before prediction.
We also implement Plain CoT Reasoner, a fine-tuned variant of Zero-shot CoT, trained to produce generic chain-of-thought reasoning.
(2) \emph{Predefined} denotes reasoning with a fixed set of manually specified perspectives; this category includes IRONIC~\cite{ramakrishnan2025ironic}, S3Agent~\cite{wang2025s3agent}, Commander-GPT~\cite{zhang2025commandergpt}.
% , and Fixed-role Reasoner that follows the three perspectives adopted from S3Agent.
% We additionally implement Fixed-role Reasoner, a fine-tuned variant trained with the three predefined perspectives from S3Agent.
We additionally implement a fine-tuned variant of S3Agent trained with the three predefined perspectives, termed Fixed-role Reasoner. 
(3) \emph{Adaptive} denotes sample-adaptive perspective generation where the analytical perspectives are self-elicited per instance.
The main table reports both zero-shot and fine-tuned systems under these settings.
All zero-shot baselines are reproduced with the same Qwen2.5-VL-7B backbone. 
All fine-tuned variants employ the same backbone and training protocol to ensure a fair comparison.
% All fine-tuned variants use the same backbone and are optimized under matched training settings for fair comparison.
% , with full details provided in Appendix~\ref{app:impl_details}.
% All zero-shot baselines are reproduced with the same Qwen2.5-VL-7B backbone, and all fine-tuned variants are first initialized by supervised fine-tuning and then further optimized with standard GRPO under the same training settings, as detailed in Appendix~\ref{app:impl_details}.

\vspace{-4pt}

% ==================================================================
%  4.2 — Main Results
% ==================================================================
\subsection{Main Results}
\label{sec:main_results}

\begin{table}[t]
\centering
% \caption{Main results on MMSD2.0, MMSD, and redEval under three reasoning paradigms. Non-trained methods use prompting only, while trained methods update model parameters. All methods use Qwen2.5-VL-7B for fair comparison. F1 is the primary metric, and best results are \textbf{bolded}.}

\caption{Main results on MMSD2.0, MMSD, and redEval. The \emph{Perspective} column indicates the reasoning paradigm: no explicit perspectives (None), predefined perspectives (Predefined), or sample-adaptive perspectives (Adaptive). The best and second-best results are in \textbf{bold} and \underline{underlined}, respectively. \ding{72} marks the best result within the zero-shot group. $^\dagger$ denotes the draft-only output.}

\label{tab:main_results}
\setlength{\tabcolsep}{3.8pt} 
\scriptsize
\begin{tabular}{l l cccc cccc cccc} 
\toprule
\textbf{Method} & \textbf{Perspective} & \multicolumn{4}{c}{\textbf{MMSD2.0}} & \multicolumn{4}{c}{\textbf{MMSD}} & \multicolumn{4}{c}{\textbf{redEval}} \\
\cmidrule(lr){3-6} \cmidrule(lr){7-10} \cmidrule(lr){11-14}
& & F1 & Acc & P & R & F1 & Acc & P & R & F1 & Acc & P & R \\
\midrule
\multicolumn{14}{l}{\emph{Zero-shot methods}} \\
\midrule
Zero-shot CoT~\cite{kojima2022large} & None & 61.6 & 71.1 & 71.9 & 53.9 & 63.2 & 73.2 & 68.1 & 59.0 & 70.6 & \zshotbest{78.2} & 75.1 & 66.6 \\
Automatic Prompt Engineer~\cite{zhou2023large} & None & 61.7 & 71.4 & 72.9 & 53.4 & 61.6 & 72.9 & 68.9 & 55.8 & 70.2 & \zshotbest{78.2} & 75.9 & 65.3 \\
Plan-and-Solve~\cite{wang2023planandsolve} & None & 54.1 & 69.0 & 74.4 & 42.5 & 57.1 & 71.9 & 70.6 & 48.0 & 58.9 & 72.0 & 77.4 & 47.6 \\
Generated Knowledge Prompting~\cite{liu2022generated} & None & 56.4 & 70.2 & \zshotbest{{76.1}} & 44.7 & 57.8 & 72.6 & \zshotbest{\textbf{72.2}} & 48.1 & 61.5 & 75.4 & \underline{80.1} & 49.9 \\
IRONIC~\cite{ramakrishnan2025ironic} & Predefined & 40.8 & 64.5 & 72.5 & 28.4 & 41.8 & 66.8 & \underline{71.4} & 29.6 & 59.1 & 76.0 & \zshotbest{\textbf{89.7}} & 44.1 \\
S3Agent~\cite{wang2025s3agent} & Predefined & 61.2 & 70.7 & 71.2 & 53.6 & 63.7 & \zshotbest{73.9} & 69.6 & 58.7 & 61.0 & 73.8 & 73.6 & 52.2 \\
Commander-GPT~\cite{zhang2025commandergpt} & Predefined & 63.3 & 65.8 & 58.8 & \zshotbest{68.6} & 61.6 & 66.0 & 55.0 & \zshotbest{70.1} & 70.1 & 72.3 & 60.9 & \zshotbest{82.5} \\
\rowcolor{gray!10}
ProCrit & Adaptive & \zshotbest{67.4} & \zshotbest{71.6} & 66.7 & 68.2 & \zshotbest{66.7} & 70.9 & 64.5 & 69.1 & \zshotbest{71.4} & 76.8 & 69.2 & 73.6 \\
% Dynamic multi-agent & None & 71.8 & 73.4 & 66.1 & 78.7 & 69.3 & 72.1 & 60.6 & 80.9 & 78.3 & 82.2 & 75.1 & 81.8 \\
\midrule
\multicolumn{14}{l}{\emph{Fine-tuned methods}} \\
\midrule
% Plain CoT Reasoner (SFT) & None & 74.8 & 77.0 & 70.9 & 79.1 & 72.8 & 76.3 & 65.8 & 81.5 & 75.6 & 79.4 & 70.9 & 80.9 \\
Plain CoT Reasoner~\cite{kojima2022large}  & None & 79.4 & 80.3 & 72.3 & \underline{88.0} & 77.5 & 79.4 & 67.5 & \underline{90.8} & 82.6 & 84.8 & 74.9 & \underline{92.2} \\
% Fixed-role Reasoner (SFT) & Predefined & 74.2 & 73.3 & 63.4 & 89.5 & 71.9 & 71.9 & 58.9 & 92.3 & 73.4 & 74.3 & 61.9 & 90.1 \\
Fixed-role Reasoner~\cite{wang2025s3agent}  & Predefined & 79.8 & 80.8 & 73.1 & 87.9 & 78.0 & 80.1 & 68.4 & \underline{90.8} & 81.6 & 83.9 & 74.3 & 90.6 \\
% Dynamic-role Reasoner (SFT) & Adaptive & 77.7 & 79.3 & 72.4 & 83.8 & 76.2 & 78.9 & 68.1 & 86.5 & 80.8 & 83.5 & 74.3 & 88.6 \\
% \rowcolor{gray!10}
% ProCrit (Vanilla) & Adaptive & 80.8 & {82.1} & {75.2} & {87.4} & \underline{78.5} & {80.9} & {70.3} & {88.9} & \underline{83.7} & \underline{86.2} & 77.9 & 90.4 \\
\rowcolor{gray!10}
ProCrit$^\dagger$ & Adaptive & \underline{81.1} & \underline{82.5} & \underline{75.7} & 87.4 & 78.4 & \underline{81.0} & 70.4 & 88.4 & \textbf{84.8} & \textbf{87.2} & 79.2 & 91.4 \\
\rowcolor{gray!10}
ProCrit & Adaptive & \textbf{83.1} & \textbf{84.0} & \textbf{76.1} & \textbf{91.5} & \textbf{80.7} & \textbf{82.6} & 71.2 & \textbf{93.2} & \underline{83.7} & \underline{85.7} & 75.7 & \textbf{93.7} \\
\bottomrule
\end{tabular}
%%}
\vspace{-10pt}
\end{table}

As shown in Table~\ref{tab:main_results}, ProCrit achieves the strongest overall performance across all three benchmarks under both zero-shot and fine-tuned settings.
%First, adaptive perspectives consistently outperform none and predefined alternatives.
First, adaptive perspective generation consistently outperforms both none and predefined alternatives.
% \paragraph{Adaptive perspectives consistently outperform none and predefined alternatives.}
% Among zero-shot methods, ProCrit~(Zero-shot) achieves the highest F1 across all three benchmarks, reaching 67.4 on MMSD2.0, 66.7 on MMSD, and 71.4 on redEval, consistently outperforming both perspective-free and predefined-perspective prompting methods. And it is noted that adaptive perspective generation yields more balanced precision--recall trade-offs by selecting analytical angles tailored to each sample. The same pattern holds under fine-tuning. Fixed-role Reasoner achieves 79.8 at F1 on MMSD2.0, slightly outperforming Plain CoT Reasoner at 79.4, which indicates that structured multi-perspective reasoning provides a marginal advantage even when perspectives are predefined. ProCrit~(Draft), which uses adaptive perspectives, further improves F1 to 81.1, confirming that sample-adaptive perspective generation yields stronger reasoning than both unstructured chain-of-thought and fixed role decomposition.
In the zero-shot regime, ProCrit attains the highest F1 score on all three benchmarks, surpassing methods that use no perspectives or predefined ones. This gain stems from its ability to select sample-specific analytical perspectives, leading to more balanced precision–recall trade-offs. The same trend holds under fine-tuning: the Fixed-Role Reasoner (79.8 F1) slightly outperforms the Plain CoT Reasoner (79.4 F1) on MMSD2.0, confirming that even predefined multi-perspective reasoning offers a modest advantage over unstructured chain-of-thought. However, ProCrit$^\dagger$, which employs adaptive perspectives, further improves F1 to 81.1, demonstrating that sample-adaptive perspective generation yields superior reasoning compared to both plain CoT and fixed-role decomposition.
Second, critic-guided revision further improves performance beyond adaptive reasoning alone.
% \paragraph{Critic-guided revision further improves performance beyond adaptive reasoning alone.}
% ProCrit~(Revise) achieves the best F1 on MMSD2.0 and MMSD, reaching 83.1 and 80.7 respectively, improving over ProCrit~(Draft) by 2.0 and 2.3 points. These gains are accompanied by improvements across all metrics on both benchmarks, demonstrating that the critic agent's targeted feedback strengthens the overall reasoning quality rather than merely shifting the precision--recall balance. A particularly notable effect is on recall: revision raises recall from 87.4 to 91.5 on MMSD2.0 and from 88.4 to 93.2 on MMSD, indicating that the critic agent helps the proposal agent recover sarcastic instances whose decisive cues are missed or under-analyzed in the initial draft.
ProCrit consistently outperforms ProCrit$^\dagger$, indicating that targeted feedback from the critic agent (ProCrit) enhances overall reasoning quality—not merely by adjusting the precision–recall balance; 
notably, recall improves substantially: on MMSD2.0, recall rises from 87.4 to 91.5, and on MMSD from 88.4 to 93.2, suggesting that the critic in (ProCrit) helps recover sarcastic instances whose decisive cues are missed or under-analyzed in the initial proposal (ProCrit$^\dagger$).
% On redEval, ProCrit~(Draft) retains the best F1 of 84.8, while ProCrit~(Revise) achieves the highest recall of 93.7. Since redEval contains a higher proportion of non-sarcastic examples with a class ratio of roughly 3:2, the additional true positives recovered by revision are offset by a precision decrease, resulting in a slightly lower F1.Nevertheless, the consistent recall improvement across all three benchmarks confirms that critic-guided revision enables the model to capture sarcastic signals that the draft reasoning alone overlooks.
On redEval, ProCrit$^\dagger$ achieves the highest F1 (84.8), while ProCrit attains the best recall (93.7). Given that redEval contains a higher proportion of non-sarcastic examples (class ratio $\approx$ 3:2), the additional true positives recovered through revision are partially offset by a drop in precision, resulting in a marginally lower F1. 
Nevertheless, the consistent recall improvement across all three benchmarks confirms that critic-guided revision (ProCrit) enables the model to capture subtle sarcastic signals overlooked by draft-only reasoning (ProCrit$^\dagger$).

\vspace{-4pt}

% ==================================================================
%  4.3 — Ablation Study
% ==================================================================
\subsection{Ablation Study}
\label{sec:ablation}

We conduct systematic ablations on MMSD2.0 and MMSD to isolate the contribution of each component.
All ablations use Qwen2.5-VL-7B as the backbone.

\textbf{Ablation on Training Stages.}
\label{sec:pipeline_ablation}
Table~\ref{tab:pipeline_ablation} analyzes the contribution of each training stage in the proposal agent.
The full training pipeline—combining SFT and RL—consistently achieves the highest F1 scores in both Draft and Revise modes on MMSD2.0 and MMSD.
Removing SFT results in the most significant performance drop, underscoring its critical role in establishing the proposal agent's sample-adaptive multi-perspective reasoning capability.
Removing RL also leads to a consistent decline, indicating its complementary function in refining reasoning quality through feedback-driven optimization.
Overall, SFT and RL serve complementary roles, and ablating either stage degrades performance across both generation modes.
%Table~\ref{tab:pipeline_ablation} demonstrates the contribution of each training stage of proposal agent. The full setting consistently achieves the best F1 in both Draft and Revise modes on MMSD2.0 and MMSD. Removing the SFT stage leads to the largest degradation, reducing Draft F1 from 81.1 to 77.3 on MMSD2.0 and from 78.4 to 77.0 on MMSD; the corresponding Revise F1 also decreases from 83.1 to 81.7 and from 80.7 to 80.4, respectively. This indicates that SFT is important for establishing the sample-adaptive multi-perspective reasoning behavior of the proposal agent. Removing the RL stage also causes a consistent decline, with Draft F1 dropping to 78.1 on MMSD2.0 and 75.9 on MMSD, and Revise F1 dropping to 82.3 and 80.2. The two stages serve complementary roles, and removing either one leads to consistent performance degradation across both generation modes.

% These results show that the SFT stage provides the proposal agent with a strong reasoning initialization, while the RL stage further improves both draft generation and feedback-guided revision.

% These results indicate that both stages are necessary: SFT establishes the multi-perspective reasoning pattern, and RL optimizes the policy for both draft generation and feedback-guided revision.

\begin{table}[h]
  \centering
  \begin{minipage}[t]{0.48\textwidth}
    \caption{Ablation on training stages of the proposal agent. Draft and Revise denote the two generation modes.}
    \label{tab:pipeline_ablation}
    \centering
    \fontsize{8.5pt}{\baselineskip}\selectfont
    \begin{tabular*}{\linewidth}{@{\extracolsep{\fill}} cc cc cc }
    \toprule
    \textbf{SFT} & \textbf{RL} & \multicolumn{2}{c}{\textbf{MMSD2.0 (F1)}} & \multicolumn{2}{c}{\textbf{MMSD (F1)}} \\
    \cmidrule(lr){3-4} \cmidrule(lr){5-6}
    &  & Draft & Revise & Draft & Revise \\
    \midrule
    \ding{55} & \ding{51} & 77.3 & 81.7 & 77.0 & 80.4 \\
    \ding{51} & \ding{55} & 78.1 & 82.3 & 75.9 & 80.2 \\
    \ding{51} & \ding{51} & \textbf{81.1} & \textbf{83.1} & \textbf{78.4} & \textbf{80.7} \\
    \bottomrule
    \end{tabular*}
  \end{minipage}%
  \hfill
  \begin{minipage}[t]{0.48\textwidth}
    \caption{Ablation on critic agent rewards.}
    \label{tab:critic_reward}
    \centering
    \fontsize{8.5pt}{\baselineskip}\selectfont
    \begin{tabular*}{\linewidth}{@{\extracolsep{\fill}} l c c }
    \toprule
    & \textbf{MMSD2.0 (F1)} & \textbf{MMSD (F1)} \\
    \cmidrule(lr){2-2} \cmidrule(lr){3-3}
    Setting & Revise & Revise \\
    \midrule
    \textbf{Full} & \textbf{83.1} & \textbf{80.7} \\
    \addlinespace
    w/o $r_{\mathrm{align}}$ & 82.2 & 79.6 \\
    w/o $r_{\mathrm{act}}$ & 82.6 & 79.4 \\
    w/o $r_{\mathrm{fmt}}^{\mathrm{c}}$ & 82.4 & 80.1 \\
    \bottomrule
    \end{tabular*}
  \end{minipage}
\end{table}

\begin{table}[h]
  \centering
  \begin{minipage}[t]{0.48\textwidth}
    \caption{Ablation on the Draft--Revise balance weight $\lambda$ in the proposal agent objective. Larger $\lambda$ assigns more weight to the Revise objective.}
    \label{tab:lambda}
    \centering
    \fontsize{8.5pt}{\baselineskip}\selectfont
    \begin{tabular*}{\linewidth}{@{\extracolsep{\fill}} c cc cc }
    \toprule
    & \multicolumn{2}{c}{\textbf{MMSD2.0 (F1)}} & \multicolumn{2}{c}{\textbf{MMSD (F1)}} \\
    \cmidrule(lr){2-3} \cmidrule(lr){4-5}
    $\lambda$ & Draft & Revise & Draft & Revise \\
    \midrule
    0 & 80.5 & 82.7 & 75.5 & 80.2 \\
    0.25 & 80.4 & 82.8 & 78.3 & 80.6 \\
    \textbf{0.50} & \textbf{81.1} & \textbf{83.1} & 78.4 & \textbf{80.7} \\
    0.75 & 80.9 & 82.4 & \textbf{78.5} & 80.1 \\
    1 & 78.4 & 82.5 & 77.8 & 79.9 \\
    \bottomrule
    \end{tabular*}
  \end{minipage}%
  \hfill
  \begin{minipage}[t]{0.48\textwidth}
    \caption{Ablation on proposal agent rewards.}
    \label{tab:proposal_reward}
    \centering
    \fontsize{8.5pt}{\baselineskip}\selectfont
    \begin{tabular*}{\linewidth}{@{\extracolsep{\fill}} l cc cc }
    \toprule
    & \multicolumn{2}{c}{\textbf{MMSD2.0 (F1)}} & \multicolumn{2}{c}{\textbf{MMSD (F1)}} \\
    \cmidrule(lr){2-3} \cmidrule(lr){4-5}
    Setting & Draft & Revise & Draft & Revise \\
    \midrule
    \textbf{Full} & \textbf{81.1} & \textbf{83.1} & 78.4 & \textbf{80.7} \\
    % \addlinespace
    % \multicolumn{5}{@{}l}{\emph{Draft-stage rewards}} \\
    w/o $r_{\mathrm{eval}}$ & 79.9 & 82.4 & 77.8 & 80.3 \\
    w/o $r_{\mathrm{acc}}^{\mathrm{d}}$ & 80.2 & 82.4 & 77.6 & 80.1 \\
    w/o $r_{\mathrm{fmt}}^{\mathrm{d}}$ & 80.5 & 82.5 & 77.7 & 80.0 \\
    % \addlinespace
    % \multicolumn{5}{@{}l}{\emph{Revision-stage rewards}} \\
    w/o $r_{\mathrm{imp}}$ & 80.7 & 82.6 & \textbf{79.0} & 79.7 \\
    w/o $r_{\mathrm{acc}}^{\mathrm{r}}$ & 80.9 & 82.2 & 78.1 & 80.2 \\
    w/o $r_{\mathrm{fmt}}^{\mathrm{r}}$ & 79.9 & 82.7 & 77.9 & 80.2 \\
    \bottomrule
    \end{tabular*}
  \end{minipage}
  \vspace{-10pt}
\end{table}

% ------------------------------------------------------------------
%  4.3.4 — Hyperparameter Analysis: λ
% ------------------------------------------------------------------
\textbf{Ablation on Draft--Revise Balance Weight $\lambda$.}
\label{sec:lambda_analysis}
% Table~\ref{tab:lambda} analyzes how the proposal agent balances Draft and Revise optimization.
% Table~\ref{tab:lambda} suggests that the proposal agent benefits from a balanced optimization of Draft and Revise, as overemphasizing either side leads to weaker overall performance. When $\lambda=0$, the objective only optimizes Draft completions. This setting yields competitive Draft F1 on MMSD2.0, but Revise performance remains below the balanced setting, showing that feedback based revision requires direct optimization rather than relying solely on a strong initial draft. When $\lambda=1$, the objective only optimizes Revise completions. This boundary weakens Draft performance, especially on MMSD2.0 where Draft F1 drops to 78.4, indicating that focusing entirely on revision reduces the quality of the initial reasoning. Intermediate values provide a better tradeoff between the two modes. The best overall performance is obtained at $\lambda=0.5$, which reaches 83.1 Revise F1 on MMSD2.0 and 80.7 on MMSD while maintaining strongest Draft performance.
% These results support the dual stage design in \S\ref{sec:proposal_rl}: Draft and Revise should be optimized jointly, with neither objective dominating the shared proposal agent.
% This indicates that Draft and Revise capture complementary capabilities, both of which are necessary for the proposal agent.
% This confirms the necessity of jointly optimizing both stages.
% : the draft objective ensures high-quality initial reasoning, while the revise objective teaches the model to effectively incorporate critic feedback.
Table~\ref{tab:lambda} presents the sensitivity of performance to the trade-off parameter $\lambda$ in Eq.~\ref{eq-10}.
The proposal agent achieves optimal results when both Draft and Revise modes are jointly optimized, with the best F1 score attained at $\lambda=0.5$.
In contrast, extreme settings degrade performance: $\lambda=0$ (optimizing only the Draft mode) and $\lambda=1$ (optimizing only the Revise mode) both yield weaker overall results.
This demonstrates that balancing the two objectives is essential—overemphasizing either stage compromises the agent's ability to generate high-quality initial proposals and effectively refine them.

\textbf{Ablation on Reward Components.}
\label{sec:reward_ablation}
(1) \emph{Proposal agent rewards.}
% We ablate rewards at both the draft and revision stages. As shown in Table~\ref{tab:proposal_reward}, removing the critic evaluation reward ($r_{\mathrm{eval}}$) causes the largest draft-stage degradation, reducing Draft F1 from 81.1 to 79.9 on MMSD2.0, confirming that the critic's quality signal provides finer-grained differentiation beyond binary accuracy. Removing $r_{\mathrm{acc}}^{\mathrm{draft}}$ and $r_{\mathrm{fmt}}^{\mathrm{draft}}$ also causes consistent degradation. For revision-stage rewards, removing $r_{\mathrm{imp}}$ degrades Revise F1 from 83.1 to 82.6 on MMSD2.0 and from 80.7 to 79.7 on MMSD, indicating that explicitly rewarding successful correction is important for effective revision.
We ablate the reward components in both Draft and Revise modes. As shown in Table~\ref{tab:proposal_reward}, in the Draft mode (Eq.~\ref{eq:draft_reward}), removing the critic evaluation reward $r_{\mathrm{eval}}$ leads to the largest performance drop, confirming that the critic's quality signal offers finer-grained supervision beyond binary accuracy. Ablating either the draft accuracy reward $r_{\mathrm{acc}}^{\mathrm{d}}$ or $r_{\mathrm{fmt}}^{\mathrm{d}}$ also results in consistent degradation.
In the Revise mode (Eq.~\ref{eq:react_reward}), removing $r_{\mathrm{imp}}$ reduces revise-stage F1 on MMSD, demonstrating that explicitly rewarding successful corrections is crucial for effective revision.
(2) \emph{Critic agent rewards.}
% As shown in Table~\ref{tab:critic_reward}, both $r_{\mathrm{align}}$ and $r_{\mathrm{act}}$ contribute meaningfully: removing $r_{\mathrm{align}}$ drops Revise F1 from 83.1 to 82.2, and removing $r_{\mathrm{act}}$ drops it to 82.6.
% The former aligns scoring with actual reasoning quality; the latter ensures feedback is actionable for revision.
% Both $r_{\mathrm{align}}$ and $r_{\mathrm{act}}$ contribute meaningfully
%The former aligns the critic's scoring with actual reasoning quality, and the latter ensures that the generated feedback provides actionable guidance for revision.
As shown in Table~\ref{tab:critic_reward}, both $r_{\mathrm{align}}$ and $r_{\mathrm{act}}$ are essential. Removing $r_{\mathrm{align}}$ decreases Revise F1 from 83.1 to 82.2, while removing $r_{\mathrm{act}}$ drops it to 82.6. The former ensures the critic's scores correlate with actual reasoning quality; the latter encourages feedback that provides actionable guidance for the proposal agent. Additionally, ablating $r_{\mathrm{fmt}}^{c}$ also harms performance. The best results are achieved when all three critic rewards are used jointly, highlighting their complementary roles.
% ------------------------------------------------------------------
%  4.3.2 — Revision Source Ablation
% ------------------------------------------------------------------

\textbf{Ablation on Revise Mode}
\label{sec:revision_ablation}
Table~\ref{tab:revision_ablation} compares four revise modes with the proposal agent fixed.
%Self-revision, where the proposal agent re-generates without external feedback, performs substantially worse than the initial draft reasoning with no revision, dropping from 81.1 to 75.9 F1 on MMSD2.0 and from 78.4 to 77.1 on MMSD.This confirms that unguided regeneration cannot reliably improve the draft reasoning; without diagnostic signals, a second attempt is more likely to disrupt correct reasoning than to fix errors.Providing external critique from a independent critic agent consistently improves over both settings above.The initialized critic, obtained by supervised distillation from teacher annotations, already raises F1 from 81.1 to 81.5 on MMSD2.0 and from 78.4 to 79.4 on MMSD, demonstrating the value of targeted natural-language feedback.

\begin{wraptable}{r}{0.52\textwidth}
\vspace{-10pt}
\centering
\caption{Ablation on the Revise mode. The proposal agent is held fixed across all experiments.}
\label{tab:revision_ablation}
\small
\vspace{6pt}
\begin{tabular}{l cc cc}
\toprule
\textbf{Revision Setting} & \multicolumn{2}{c}{\textbf{MMSD2.0}} & \multicolumn{2}{c}{\textbf{MMSD}} \\
\cmidrule(lr){2-3} \cmidrule(lr){4-5}
& F1 & Acc & F1 & Acc \\
\midrule
None & 81.1 & 82.5 & 78.4 & 81.0 \\
Self-revision & 75.9 & 79.0 & 77.1 & 80.2 \\
Initialized critic & 81.5 & 82.9 & 79.4 & 81.9 \\
\textbf{Proposal-refined critic} & \textbf{83.1} & \textbf{84.0} & \textbf{80.7} & \textbf{82.6} \\
\bottomrule
\end{tabular}
\vspace{-10pt}
\end{wraptable}

Self-revision, where the proposal agent re-generates its output without external feedback, performs substantially worse than the initial draft reasoning with no revision (None). This indicates that unguided regeneration is unreliable: in the absence of diagnostic signals, a second pass is more likely to corrupt correct reasoning than to correct errors.
In contrast, incorporating external critique from an independent critic agent consistently improves performance over both baselines. Specifically, the initialized critic, obtained via supervised distillation from teacher annotations, yields measurable gains, highlighting the effectiveness of targeted natural-language feedback.
Further improvement is achieved through mutual-refinement reinforcement learning, where the proposal-refined critic attains the best results. This validates our co-evolutionary design: revision outcomes from the proposal agent provide outcome-grounded signals that enable the critic to transcend mere imitation and generate more actionable and context-aware guidance. 
Finally, we observe that a single revision round captures the dominant improvement (see Appendix~\ref{app:additional_ablations}).

\vspace{-4pt}
\section{Conclusion}
\label{sec:conclusion}
\vspace{-2pt}
This paper presents ProCrit, a two-agent framework that shifts multimodal sarcasm detection from relying on manually prescribed analytical perspectives to self-elicited multi-perspective reasoning.
To obtain process-level supervision for this capability, we design a dynamic-role agentic rollout that synthesizes sample-adaptive reasoning annotations where each analytical perspective progressively builds on preceding analyses.
To improve reasoning reliability, ProCrit adopts a critic-guided draft--critique--revise paradigm with a mutual-refinement training strategy that jointly optimizes reasoning and revision quality through external critique, while reciprocally refining the critic through actual revision outcomes.
Extensive experiments on MMSD2.0, MMSD, and redEval demonstrate that ProCrit achieves the strongest overall performance across all three benchmarks, and ablation studies confirm the contribution of each component.
We provide further discussion on limitations and broader impact in the Appendix~\ref{app:limitations} and~\ref{app:broader_impact}.

% \section*{References}

% References follow the acknowledgments in the camera-ready paper. Use unnumbered first-level heading for
% the references. Any choice of citation style is acceptable as long as you are
% consistent. It is permissible to reduce the font size to \verb+small+ (9 point)
% when listing the references.
% Note that the Reference section does not count towards the page limit.
% \medskip

% {
% \small

% [1] Alexander, J.A.\ \& Mozer, M.C.\ (1995) Template-based algorithms for
% connectionist rule extraction. In G.\ Tesauro, D.S.\ Touretzky and T.K.\ Leen
% (eds.), {\it Advances in Neural Information Processing Systems 7},
% pp.\ 609--616. Cambridge, MA: MIT Press.

% [2] Bower, J.M.\ \& Beeman, D.\ (1995) {\it The Book of GENESIS: Exploring
%   Realistic Neural Models with the GEneral NEural SImulation System.}  New York:
% TELOS/Springer--Verlag.

% [3] Hasselmo, M.E., Schnell, E.\ \& Barkai, E.\ (1995) Dynamics of learning and
% recall at excitatory recurrent synapses and cholinergic modulation in rat
% hippocampal region CA3. {\it Journal of Neuroscience} {\bf 15}(7):5249-5262.
% }

\newpage

% {
% \small
% \bibliographystyle{ieeenat_fullname}
% \bibliography{local-bib}
% }

{
\small

\bibliographystyle{IEEEtran}
\bibliography{local-bib}
}

%%%%%%%%%%%%%%%%%%%%%%%%%%%%%%%%%%%%%%%%%%%%%%%%%%%%%%%%%%%%

%%%%%%%%%%%%%%%%%%%%%%%%%%%%%%%%%%%%%%%%%%%%%%%%%%%%%%%%%%%%

\endgroup

\appendix

\clearpage
\setcounter{page}{1}
\setcounter{table}{5}
\setcounter{figure}{5}

\begin{center}
    \fontsize{20pt}{\baselineskip}\selectfont
    \textbf{Appendix}
    \vspace{0.4cm}
\end{center}

\tableofcontents
\newpage

\section{Related Work}
\label{sec:related}

\subsection{Multimodal Sarcasm Detection}

% 现有方法全部是端到端分类（输入→标签），不关心推理过程。即使用了 LLM，也只是把 LLM 当特征提取器或 prompt 工具。没有工作让模型产出可解释的推理链，更没有对推理过程进行评估和修正。

Early multimodal sarcasm detection methods treat the task as end-to-end classification, fusing cross-modal features through attention mechanisms~\cite{cai2019multi}, graph convolutional networks~\cite{liang2022multi}, contrastive learning~\cite{liang2024fusion}, or external knowledge injection~\cite{yue2023knowledgenet}, and mapping the fused representation directly to a sarcasm label.
More recent variants incorporate LLMs as feature enhancers via demonstration retrieval~\cite{tang2024leveraging} or debate-guided text generation~\cite{zhou2025ldgnet}.
While these methods have steadily improved detection accuracy, their analytical process remains implicit in the learned representations, with no explicit reasoning over the input.
A growing line of work makes the analytical process explicit by prompting LLMs and VLMs to produce intermediate reasoning alongside predictions.
S3~Agent~\cite{wang2025s3agent} assigns three fixed analytical perspectives and Commander-GPT~\cite{zhang2025commandergpt} routes inputs through six sub-task agents, each aggregated by a decision module, while IRONIC~\cite{ramakrishnan2025ironic} predicts predefined discourse relations to guide interpretive reasoning.
However, these methods rely on predefined perspectives that operate independently, limiting adaptability to diverse sarcasm mechanisms and preventing later analyses from building on earlier findings.
Our work instead realizes \emph{self-elicited multi-perspective reasoning}, where the model dynamically generates sample-adaptive perspectives and progressively integrates them into a coherent analysis, with an independent critic agent providing external evaluation and targeted revision guidance.

% 角色预定义，而讽刺机制不同需要的分析角度不同 / 无跨视角交互共享上下文 / 无训练、评估修正

% A growing line of work moves beyond direct classification by prompting LLMs and VLMs to produce analytical reasoning alongside predictions. S3~Agent~\cite{wang2025s3agent} assigns three fixed perspectives (expression, semantics, sentiment) and Commander-GPT~\cite{zhang2025commandergpt} routes inputs through six sub-task agents (sentiment, rhetoric, facial expression, etc.), each aggregated by a decision module.
% IRONIC~\cite{ramakrishnan2025ironic} takes a different angle by predicting predefined discourse coherence relations and using them to guide interpretive reasoning.
% Despite generating intermediate analyses, these methods share three fundamental limitations:(1)~analytical perspectives or agent roles are \emph{predefined} by the designer, limiting adaptability to the diverse mechanisms through which sarcasm manifests;
% (2)~agents analyze independently with \emph{no cross-perspective interaction}, so later analyses cannot build on earlier findings;
% and (3)~reasoning quality is bounded by the backbone model's prompting ability, with \emph{no task-specific training to internalize the nuanced mechanisms of sarcasm} and \emph{no mechanism for evaluating or correcting} the generated output.
% Our work addresses all three: we dynamically generate context-aware analytical roles (\S\ref{sec:data_synthesis}), train a dedicated critic to assess reasoning quality and provide actionable feedback (\S\ref{sec:critic_rl}), and optimize both draft and revision capabilities through reinforcement learning (\S\ref{sec:proposal_rl}).

\subsection{Reasoning and Self-Refinement}
Chain-of-thought prompting~\cite{wei2022chain} and its extensions---Tree-of-Thoughts~\cite{yao2023tree}, self-consistency decoding~\cite{wang2023selfconsistency}, and multi-agent debate~\cite{du2024improving,liang2024encouraging}---have established that explicit step-by-step reasoning and diverse-path exploration substantially improve LLM performance.
A related line of work explores \emph{self-refinement}: Self-Refine~\cite{madaan2023selfrefine} and Reflexion~\cite{shinn2023reflexion} demonstrate that iterative generate-critique-revise loops can improve output quality at inference time.
% However, Huang et al.~\cite{huang2024large} show that without external feedback, LLMs often degrade rather than improve their answers---motivating the use of a \emph{separately trained} critic rather than self-evaluation.
However, a growing body of evidence~\cite{huang2024large,kamoi2024when,tsui2025selfcorrection,zhan2026selfcorrect,li2026decomposing} indicates that \emph{intrinsic} self-correction without external signals remains fundamentally unreliable---motivating the use of a \emph{independent} critic rather than self-evaluation.
% Existing critic models provide scalar quality signals but not revision guidance: process reward models~\cite{lightman2024lets} and Math-Shepherd~\cite{wang2024mathshepherd} score reasoning steps, and Critic-V~\cite{zhang2024criticv} extends this to VLM reasoning, yet none produces natural language feedback.
% In contrast, our framework couples calibrated scoring with actionable natural-language feedback
% % Our critic agent differs by generating both calibrated scores and actionable feedback
% , with feedback quality explicitly optimized via RL to improve the proposal agent's output (\S\ref{sec:critic_rl}).
Existing critic models provide scalar quality signals but not revision guidance: process reward models~\cite{lightman2024lets} and Math-Shepherd~\cite{wang2024mathshepherd} score reasoning steps, and Critic-V~\cite{zhang2024criticv} extends this to VLM reasoning, yet none produces natural-language feedback.
Our critic agent generates both calibrated scores and actionable feedback, with feedback quality explicitly optimized via RL to improve the proposal agent's revision outcomes.

\subsection{Reinforcement Learning for LLM Reasoning}
RLHF~\cite{ouyang2022training} is the standard post-training alignment paradigm.
For reasoning, Group Relative Policy Optimization (GRPO)~\cite{shao2024deepseekmath} eliminates the value model by computing group-relative advantages, and has been scaled in DeepSeek-R1~\cite{deepseek2025r1} and Open-Reasoner-Zero~\cite{chen2025openreasonerzero}.
Zhai et al.~\cite{zhai2025sft} show that SFT stabilizes format while RL drives generalization, supporting SFT-then-RL pipelines.
Recent work extends GRPO-style training to vision-language models~\cite{vlmr1,visionr1} for multimodal reasoning.
However, existing RL approaches optimize single-pass generation with no opportunity to revise, and rely on outcome-based or scalar process rewards without natural-language feedback.
Our dual-stage GRPO (\S\ref{sec:proposal_rl}) jointly optimizes draft generation and feedback-driven revision within a single model, with a frozen critic supplying both scalar scores and natural-language feedback during training.

% RLHF~\cite{ouyang2022training} is the standard post-training alignment paradigm.
% For reasoning, Group Relative Policy Optimization (GRPO)~\cite{shao2024deepseekmath} eliminates the value model by computing group-relative advantages, and has been scaled in DeepSeek-R1~\cite{deepseek2025r1} and Open-Reasoner-Zero~\cite{chen2025openreasonerzero}.
% Zhai et al.~\cite{zhai2025sft} show that SFT stabilizes format while RL drives generalization, providing the rationale for SFT-then-RL pipelines.
% Recent work extends GRPO-style training to vision-language models~\cite{vlmr1,visionr1} for multimodal reasoning.
% However, existing RL approaches for reasoning share two characteristics that our work departs from: they optimize \emph{single-pass} generation with no opportunity to revise, and reward signals are outcome-based or scalar process scores with no use of natural language feedback.
% Our dual-stage GRPO (\S\ref{sec:proposal_rl}) jointly optimizes first-pass drafting and feedback-driven revision within a single model, with a frozen critic supplying both scalar scores and natural language feedback during training.

% suggests directions for revision
% To our knowledge, this is also the first application of GRPO to multimodal sarcasm detection---a subjective understanding task that differs substantially from the mathematical and coding domains where RL for reasoning has been explored.

%% ================================================================
%%  APPENDIX A — Process-Level Reasoning Annotation Synthesis
%% ================================================================
\section{Process-Level Reasoning Annotation Synthesis}
% \label{app:data_construction}
\label{app:data_synthesis}

% This section details the process-level reasoning annotation synthesis pipeline underlying ProCrit. We first discuss limitations of existing annotations that motivate our approach (\S\ref{app:annotation_comparison}), then describe the dynamic-role rollout protocol and the resulting data characteristics (\S\ref{app:rollout_protocol}), including quantitative analyses of perspective distributions (\S\ref{app:data_analysis}), and finally provide concrete synthesis examples (\S\ref{app:synthesis_example}).

\begin{figure}[h]
    \centering
    \includegraphics[width=1\linewidth]{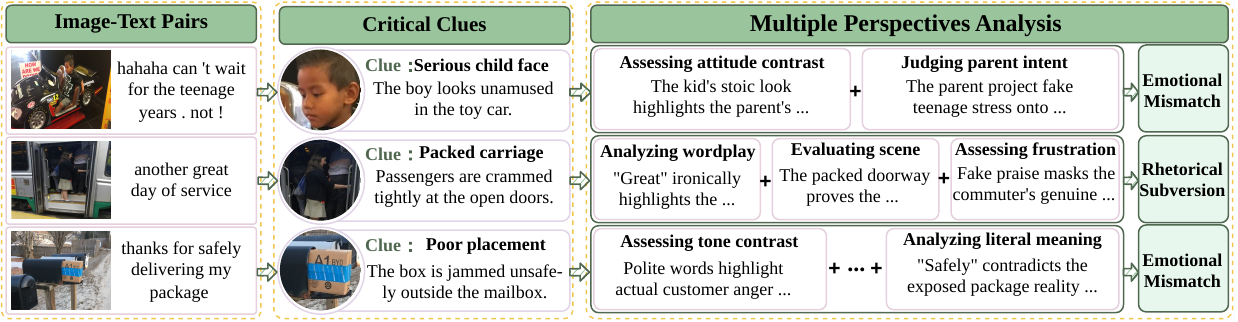}
    \caption{Examples illustrating that different sarcastic samples involve distinct mechanisms and demand different combinations of analytical perspectives to expose the text--image incongruity.}
    \label{fig:perspective_example}
\end{figure}

This section details the process-level reasoning annotation synthesis pipeline underlying ProCrit. 
As different sarcastic samples rely on distinct mechanisms and require different analytical perspectives to expose the underlying incongruity (see Fig.~\ref{fig:perspective_example}), our synthesis pipeline generates sample-adaptive, multi-perspective reasoning annotations. 
We describe the dynamic-role rollout protocol (\S\ref{app:rollout_protocol}), present quantitative analyses of the synthesized data (\S\ref{app:data_analysis}), and provide concrete synthesis examples (\S\ref{app:synthesis_example}).

\begin{figure}[h]
    \centering
    \includegraphics[width=1\linewidth]{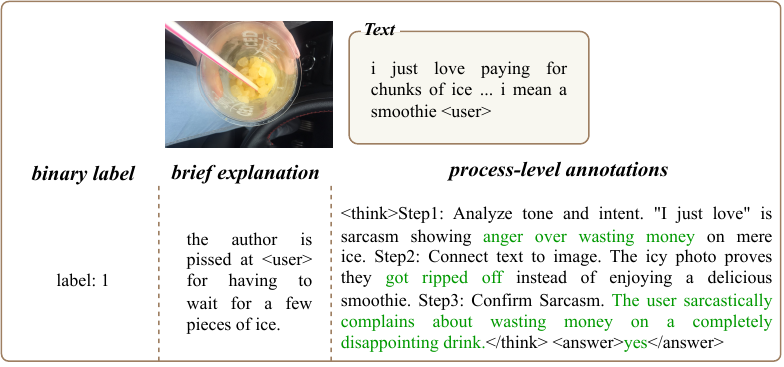}
    \caption{Comparison between existing dataset annotations and the process-level reasoning annotations synthesized by ProCrit.}
    \label{fig:comparison}
\end{figure}

\subsection{Dynamic-Role Rollout Details}
\label{app:rollout_protocol}

% While existing sarcasm corpora~\cite{qin2023mmsd2,desai2022nice} offer binary labels or brief explanations of the intended meaning, they lack process-level annotations that capture the reasoning steps or critical cues through which sarcastic intent is identified. 
% To bridge this gap, the dynamic-role agentic rollout (\S\ref{sec:data_synthesis}) uses GLM-4.6V~\cite{glm2024chatglm} as the teacher model~$\mathcal{T}$ to synthesize multi-perspective, context-aware reasoning annotations. 
% Figure~\ref{fig:comparison} illustrates the comparison between existing annotations and our synthesized process-level reasoning annotations.
% The dynamic-role agentic rollout (\S\ref{sec:data_synthesis}) uses GLM-4.6V~\cite{glm2024chatglm} as the teacher model~$\mathcal{T}$.
% The source image-text pairs are drawn from the MMSD2.0 training set, which contains 19,816 samples (40.8\% sarcastic).
% For each training example, we stochastically sample $N{=}2$ trajectories with temperature 0.7.
While existing sarcasm corpora~\cite{qin2023mmsd2,desai2022nice} offer binary labels or brief explanations of the intended meaning, they lack process-level annotations that capture the reasoning steps or critical cues through which sarcastic intent is identified.
As illustrated in Figure~\ref{fig:comparison}, the dynamic-role agentic rollout (\S\ref{sec:data_synthesis}) addresses this by synthesizing multi-perspective, context-aware reasoning annotations from raw image-text samples.
Specifically, we use GLM-4.6V~\cite{glm2024chatglm} as the teacher model~$\mathcal{T}$ and draw source samples from the MMSD2.0 training set.
% Concretely, at the first turn, $\mathcal{T}$ receives an initialization prompt (Prompt~1) that specifies the structured JSON output format for each reasoning step; at every subsequent turn, a follow-up prompt (Prompt~2) elicits a new analytical perspective conditioned on the accumulated analyses, following the rollout procedure of Eq.~\ref{eq:rollout}.
Trajectories whose final answers are correct are directly flattened into single-turn draft training sequences.
% For trajectories that produce incorrect answers, we collect the flawed reasoning and use the critic agent to generate diagnostic feedback (Prompt~6), which is then provided to the teacher model to re-generate a corrected analysis (Prompt~7).
For trajectories that produce incorrect answers, we collect the flawed reasoning. After the critic agent is initialized by SFT using teacher-annotated draft evaluations, we use this SFT critic to generate diagnostic feedback (Prompt~6) for trajectories with incorrect final answers. The feedback is then provided to the teacher model~$\mathcal{T}$ to regenerate corrected analyses (Prompt~7).
This produces draft--feedback--revision triples that serve as multi-turn training data for the proposal agent's revision capability.
The resulting training dataset contains 9,000 sequences: 6,780 single-turn draft sequences (from initially correct trajectories) and 2,220 draft--revision pairs (from critic-guided correction of initially incorrect trajectories).

\subsection{Synthesized Data Analysis}
\label{app:data_analysis}

We analyze the raw synthesized reasoning trajectories (15,042 valid samples before filtering) to characterize the data produced by the dynamic-role rollout.

\paragraph{Step count distribution.}
Table~\ref{tab:step_distribution} reports the distribution of reasoning steps across all synthesized trajectories. The majority of samples (73.9\%) use 3 analytical steps, while 14.8\% require 4 steps and 10.9\% use 2 steps. Only 0.3\% of samples use 5 steps. Notably, sarcastic samples tend to require more reasoning steps on average (3.12 vs.\ 2.95 for non-sarcastic), reflecting the greater analytical complexity involved in identifying cross-modal incongruities. Sarcastic samples also exhibit a higher proportion of 4-step trajectories (18.3\% vs.\ 10.6\%), suggesting that uncovering sarcastic intent often demands additional analytical perspectives beyond the baseline.

\begin{table}[h]
\centering
\caption{Reasoning step count distribution across all synthesized trajectories, broken down by sarcasm label.}
\label{tab:step_distribution}
\small
\begin{tabular}{l rrrr|c}
\toprule
 & \multicolumn{1}{c}{\textbf{2-step}} & \multicolumn{1}{c}{\textbf{3-step}} & \multicolumn{1}{c}{\textbf{4-step}} & \multicolumn{1}{c}{\textbf{5-step}} & \textbf{Avg.} \\
\midrule
Sarcastic     & 578 (7.1\%) & 6,027 (74.1\%) & 1,491 (18.3\%) & 38 (0.5\%) & 3.12 \\
Non-sarcastic & 1,069 (15.5\%) & 5,096 (73.8\%) & 733 (10.6\%) & 10 (0.1\%) & 2.95 \\
\midrule
All           & 1,647 (10.9\%) & 11,123 (73.9\%) & 2,224 (14.8\%) & 48 (0.3\%) & 3.05 \\
\bottomrule
\end{tabular}
\end{table}

% \paragraph{Perspective distribution.}
% We categorize the 41,721 step titles extracted from all trajectories into semantic perspective types. As shown in Table~\ref{tab:perspective_distribution}, the most frequently adopted perspectives are \emph{Text/Caption Analysis} (45.5\%) and \emph{Visual/Image Content Analysis} (20.5\%), reflecting the fundamental need to interpret both modalities. \emph{Sarcasm/Irony Judgment} steps (23.8\%) typically appear as the final step, integrating findings from prior analyses. Notably, \emph{Cross-modal Contrast/Mismatch} (3.0\%) and \emph{Tone/Sentiment Analysis} (2.8\%) emerge as specialized perspectives invoked when the standard visual--textual analysis is insufficient, demonstrating the model's ability to adaptively select perspectives based on sample characteristics.

% \begin{table}[h]
% \centering
% \caption{Distribution of analytical perspective types across all synthesized reasoning steps.}
% \label{tab:perspective_distribution}
% \small
% \begin{tabular}{lcc}
% \toprule
% \textbf{Perspective Category} & \textbf{Count} & \textbf{Proportion} \\
% \midrule
% Text/Caption Analysis & 18,974 & 45.5\% \\
% Sarcasm/Irony Judgment & 9,936 & 23.8\% \\
% Visual/Image Content Analysis & 8,568 & 20.5\% \\
% Cross-modal Contrast/Mismatch & 1,266 & 3.0\% \\
% Tone/Sentiment Analysis & 1,171 & 2.8\% \\
% Contextual/Cultural Analysis & 60 & 0.1\% \\
% Other & 1,746 & 4.2\% \\
% \bottomrule
% \end{tabular}
% \end{table}

\paragraph{Label distribution.}
Among the 15,042 raw synthesized trajectories, 8,134 (54.1\%) are labeled as sarcastic and 6,908 (45.9\%) as non-sarcastic, reflecting a roughly balanced distribution consistent with the source MMSD2.0 training set.

\subsection{Synthesis Examples}
\label{app:synthesis_example}

\begin{figure}[htbp]
    \centering
    \includegraphics[width=1\linewidth]{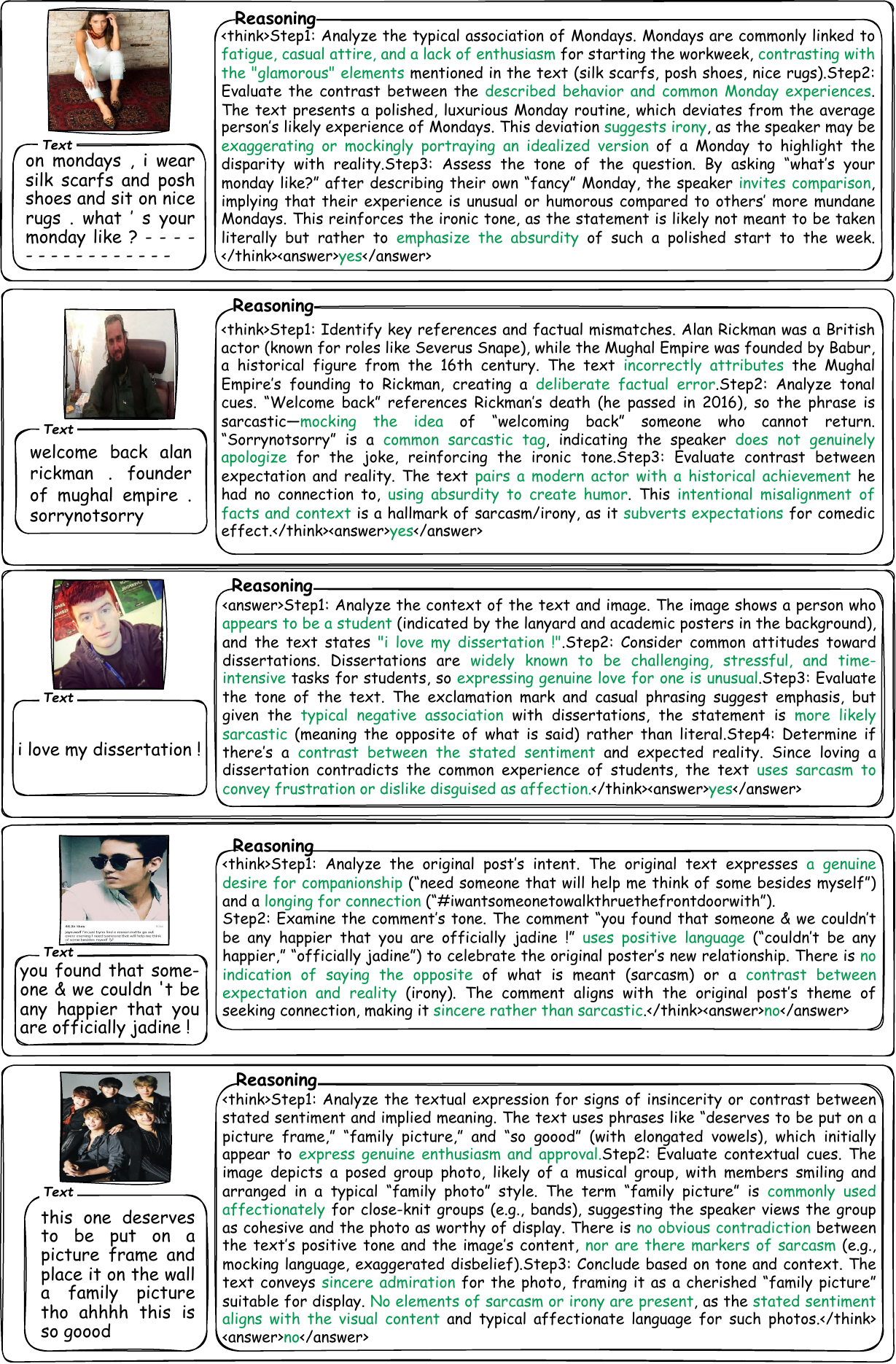}
    \caption{Examples of reasoning trajectory synthesized through dynamic-role agentic rollout. Each analytical perspective is spawned sequentially within a shared context, progressively building a coherent multi-perspective analysis.}
    \label{fig:synthetic_data}
\end{figure}

Figures~\ref{fig:synthetic_data} present concrete examples of the process-level reasoning annotations produced by the dynamic-role agentic rollout, demonstrating the sample-adaptive selection of perspectives based on sample characteristics.

%% ================================================================
%%  APPENDIX B — Training Details
%% ================================================================
\section{Training Details}
\label{app:training_details}
\label{app:impl_details}

This section reports the full implementation and training configuration for ProCrit.

\subsection{Hyperparameters}
\label{app:hyperparams}
We have illustrated the key hyper-parameters in \S\ref{sec:setup}.
In this section, we provide more information
about the hyper-parameters used in our experiment. 
Tables~\ref{tab:hyper_sft} and~\ref{tab:hyper_grpo} summarize the specific hyperparameter settings for supervised fine-tuning and reinforcement learning, respectively. For the dual-stage GRPO, we utilize the OpenR1 framework for training.

\begin{table}[h]
\centering
\begin{minipage}[b][4cm][t]{0.48\textwidth}
\centering
\caption{Hyperparameters for supervised fine-tuning.}
\label{tab:hyper_sft}
\vfill 
\resizebox{\linewidth}{!}{%
\begin{tabular}[b]{lcc}
\toprule
\textbf{Hyperparameter} & \textbf{Proposal Agent} & \textbf{Critic Agent} \\
\midrule
Epochs & 3 & 3 \\
Learning rate & $1 \times 10^{-5}$ & $1 \times 10^{-5}$ \\
Warmup ratio & 0.1 & 0.1 \\
Per-device batch size & 4 & 4 \\
Gradient accumulation & 2 & 2 \\
Max sequence length & 2048 & 2048 \\
Precision & bf16 & bf16 \\
\bottomrule
\end{tabular}%
}
\end{minipage}\hfill
\begin{minipage}[b][4cm][t]{0.48\textwidth}
\centering
\caption{Hyperparameters for GRPO reinforcement learning.}
\label{tab:hyper_grpo}
\vfill
\resizebox{\linewidth}{!}{%
\begin{tabular}[b]{lcc}
\toprule
\textbf{Hyperparameter} & \textbf{Proposal Agent} & \textbf{Critic Agent} \\
\midrule
Epochs & 4 & 4 \\
Learning rate & $3 \times 10^{-6}$ & $1 \times 10^{-6}$ \\
Per-device batch size & 8 & 8 \\
Gradient accumulation & 4 & 4 \\
Max completion length & 512 & 768 \\
KL coefficient $\beta$ & 0.02 & 0.0 \\
Clip ratio $\epsilon$ & 0.2 & 0.2 \\
Precision & bf16 & bf16 \\
\bottomrule
\end{tabular}%
}
\end{minipage}
\end{table}

\subsection{Prompts}
\label{app:prompts}
This subsection provides the complete prompts used in reasoning annotation synthesis (\S\ref{sec:data_synthesis}), proposal agent inference (drafting and revision), and critic evaluation, as referenced throughout \S\ref{sec:framework}.
\paragraph{Reasoning annotation synthesis.}
\label{app:synthesis_prompt}
During the dynamic-role agentic rollout (\S\ref{sec:data_synthesis}), the teacher model~$\mathcal{T}$ receives an initialization prompt (Prompt~1) that specifies the structured JSON output format for each reasoning step.
At every subsequent turn, a follow-up prompt (Prompt~2) elicits a new analytical perspective conditioned on the accumulated analyses, following the rollout procedure of Eq.~\ref{eq:rollout}.
\paragraph{Proposal drafting.}
\label{app:draft_prompt}
At inference time, the proposal agent uses the \emph{dynamic-perspective} prompt (Prompt~3), in which the model autonomously determines the analytical perspectives needed for each sample.
For controlled comparison, we additionally design two ablation baselines: a \emph{fixed-perspective} prompt (Prompt~4) that prescribes three predefined perspectives, and a \emph{generic} prompt (Prompt~5) that provides a minimal chain-of-thought instruction without perspective guidance.
These two variants correspond to the Fixed-role Reasoner and Plain CoT Reasoner baselines in Table~\ref{tab:main_results}, respectively.

\begin{center}
\begin{tcolorbox}[colback=cyan!5!white,colframe=cyan!50!black,width=0.95\linewidth,label={box:synthesis_system},title={\textit{\textbf{Prompt 1: Data Synthesis System Prompt}}}]
{\small
You are an expert AI assistant that explains your reasoning step by step. 
For each step, provide a title that describes what you're doing in that step, along with the content. 
Decide if you need another step or if you're ready to give the final answer. 
Respond in JSON format with 'title', 'content', and 'next\_action' (either 'continue' or 'final\_answer') keys. 
USE AT LEAST 2 STEPS OF REASONING. 
Decide confidently and never respond with "not enough context", "cannot determine", or any form of hedging. YOU CAN BE WRONG. WHEN YOU SAY YOU ARE RE-EXAMINING, ACTUALLY RE-EXAMINE, AND USE ANOTHER APPROACH TO DO SO. DO NOT JUST SAY YOU ARE RE-EXAMINING. USE AT LEAST 3 METHODS TO DERIVE THE ANSWER. USE BEST PRACTICES.
Example of a valid JSON response:
\texttt{\{"title": "Identifying Key Information",
            "content": "To begin solving this problem, we need to carefully examine the given information and identify the crucial elements that will guide our solution process. This involves...",
            "next\_action": "continue"\}}
Strictly follow the JSON response format. The response should start with \texttt{\{}. DO NOT START WITH "```json" OR ANYTHING ELSE.

}
\end{tcolorbox}
\end{center}

\begin{center}
\begin{tcolorbox}[colback=cyan!5!white,colframe=cyan!50!black,width=0.95\linewidth,label={box:synthesis_followup},title={\textit{\textbf{Prompt 2: Data Synthesis Follow-up Prompt}}}]
{\small
Continue with exactly one new reasoning step.
Return a valid JSON object with keys: \texttt{title}, \texttt{content}, \texttt{next\_action}.
Please provide an one-more different step analysis here that you believe is likely to answer the question correctly.
Do not provide the final answer yet unless you have already produced at least 2 reasoning steps.
If you are not done, set \texttt{next\_action} to \texttt{continue}.
}
\end{tcolorbox}
\end{center}

\begin{center}
\begin{tcolorbox}[colback=cyan!5!white,colframe=cyan!50!black,width=0.95\linewidth,label={box:dynamic_prompt},title={\textit{\textbf{Prompt 3: Dynamic-Perspective Proposal Drafting Prompt}}}]
{\small
\textbf{\#\#\# Question}

\texttt{<image>}

Text: \{text\}

Does the composite message of this image-text pair qualify as ironic/sarcastic?

\textbf{\#\#\# Instruction}

1. Output step-by-step analysis strictly within \texttt{<think>...</think>} tags.

Requirements:
\begin{itemize}[nosep,leftmargin=1em]
\item Decide the necessary number of steps (recommended: 3--5)
\item Assign each step a clear, incisive title (e.g., ``Step2: Pragmatic intent decoding.'')
\item For each step, explicitly select an analytical perspective that is:
  \begin{itemize}[nosep,leftmargin=1em]
  \item most relevant to the characteristics of the specific image-text pair
  \item logically builds upon / responds to the findings and open questions from previous steps
  \end{itemize}
\item Provide concise, critical, evidence-focused analysis under each step (avoid speculation, prioritize observable multimodal cues)
\end{itemize}

2. After completing all steps, output the final answer within \texttt{<answer>...</answer>} tags using one of the following options only:
\begin{itemize}[nosep,leftmargin=1em]
\item yes (sarcasm or irony is present)
\item no (sarcasm or irony is not present)
\end{itemize}
}
\end{tcolorbox}
\end{center}

\begin{center}
\begin{tcolorbox}[colback=cyan!5!white,colframe=cyan!50!black,width=0.95\linewidth,label={box:fixed_prompt},title={\textit{\textbf{Prompt 4: Fixed-Perspective Proposal Drafting Prompt}}}]
{\small
\textbf{\#\#\# Question}

\texttt{<image>}

Text: \{text\}

Does the composite message of this image-text pair qualify as ironic/sarcastic?

\textbf{\#\#\# Instruction}

1. Output the analysis strictly within \texttt{<think>...</think>} tags.

Use exactly only the following three fixed perspectives, in this exact order and with these exact step titles:

\textbf{Step1: Surface-Level Discrepancy Analysis.}
Analyze whether there is an obvious mismatch, exaggeration, reversal, or unexpected contrast between the image and the text at the surface level.

\textbf{Step2: Semantic Relation Analysis.}
Analyze how the text semantically relates to the image: whether it reinforces the image, contradicts it, reframes it, mocks it, or creates an implied meaning beyond the literal content.

\textbf{Step3: Sentiment \& Tone Analysis.}
Analyze the tone, attitude, emotional polarity, and pragmatic intent of the text in relation to the image, focusing on whether the speaker's apparent attitude signals sarcasm or irony.

2. After completing the three steps, output the final answer within \texttt{<answer>...</answer>} tags using one of the following options only:
\begin{itemize}[nosep,leftmargin=1em]
\item yes (sarcasm or irony is present)
\item no (sarcasm or irony is not present)
\end{itemize}
}
\end{tcolorbox}
\end{center}

\begin{center}
\begin{tcolorbox}[colback=cyan!5!white,colframe=cyan!50!black,width=0.95\linewidth,label={box:generic_prompt},title={\textit{\textbf{Prompt 5: Generic Proposal Drafting Prompt}}}]
{\small
\textbf{\#\#\# Question}

\texttt{<image>}

Text: \{text\}

Does the composite message of this image-text pair qualify as ironic/sarcastic?

\textbf{\#\#\# Instruction}

Analyze the image-text pair with a chain of thought, then determine whether it is sarcastic.

1. Output the analysis strictly within \texttt{<think>...</think>} tags.
Detailed analysis and reasoning steps supporting the conclusion.

2. After completing the three steps, output the final answer within \texttt{<answer>...</answer>} tags using one of the following options only:
\begin{itemize}[nosep,leftmargin=1em]
\item yes (sarcasm or irony is present)
\item no (sarcasm or irony is not present)
\end{itemize}
}
\end{tcolorbox}
\end{center}

\paragraph{Critic evaluation.}
\label{app:critic_prompt}

The critic agent uses the following prompt (Prompt~6) to evaluate the quality of the proposal's reasoning process. The scoring rubric defines a 0--2 scale that evaluates interpretation accuracy, cross-modal reasoning, and reasoning coherence.

\begin{center}
\begin{tcolorbox}[colback=cyan!5!white,colframe=cyan!50!black,width=0.95\linewidth,label={box:critic_prompt},title={\textit{\textbf{Prompt 6: Critic Evaluation Prompt}}}]
{\small
You are an expert evaluator for multimodal sarcasm/irony detection reasoning.

You are given an image-text pair along with a reasoning chain and concluded answer that analyzes whether this pair conveys sarcasm/irony.

\textbf{\#\#\# Input}

Text: \{text\}

Reasoning and answer:

\{reasoning\}

\textbf{\#\#\# Your task}

Evaluate the quality of this reasoning process --- whether it correctly understood what is happening in this image-text pair and built a sound argument.

A correct conclusion does not guarantee a high score --- evaluate the reasoning process itself, not just whether the answer matches.

Focus on:
\begin{enumerate}[nosep,leftmargin=1.5em]
\item \textbf{Interpretation accuracy (primary)} --- Does the reasoning correctly interpret the combined meaning of the image-text pair and explain WHY it is or isn't sarcastic?
\item \textbf{Cross-modal reasoning} --- Does it connect image and text into joint reasoning? For sarcastic pairs: does it identify how they contradict or recontextualize each other to create irony? For non-sarcastic pairs: does it show how they reinforce the same tone and real meaning?
\item \textbf{Reasoning coherence and efficiency} --- Does the evidence chain build logically toward the conclusion, with each step contributing a concrete cue or reasoning move? Penalize unsupported leaps, contradictions, and filler steps that do not serve the final judgment.
\end{enumerate}

Rate the reasoning on a 0--2 scale.

\textbf{\#\#\# Scoring rubric}

\textbf{0 = Misunderstanding} --- the reasoning does not correctly understand this image-text pair. Common patterns:
\begin{itemize}[nosep,leftmargin=1.5em]
\item Treating any image-text contrast as automatic sarcasm
\item Confusing playful banter, humor, or jokes with sarcasm
\item Treating sarcasm-related words as proof of sarcastic intent
\item Assuming positive wording must be sincere, ignoring image contradiction
\item Missing subtle sarcasm due to a narrow definition
\item Analyzing text in isolation without considering the image
\item Treating absence of exaggeration as evidence of sincerity
\end{itemize}

\textbf{1 = Partial understanding} --- the reasoning is on the right track but the analysis is weak. Common patterns:
\begin{itemize}[nosep,leftmargin=1.5em]
\item Identifies some relevant cues but misses others
\item Stays surface-level without explaining cross-modal interaction
\item Includes filler, detours, or redundant steps
\item Has logical gaps or weak connections between steps
\end{itemize}

\textbf{2 = Strong understanding} --- The reasoning clearly interprets what this pair is trying to express, precisely pinpoints the specific elements that create the sarcastic/non-sarcastic effect, and builds on joint image-text reasoning. Key cues are not missed, and the conclusion follows naturally.

\textbf{Feedback} (2--4 sentences): Briefly note what was done well or poorly. If the reasoning is weak or wrong, point out what was missed, misread, or insufficiently supported and indicate what the reasoning should have focused on instead.

\textbf{\#\#\# Output format}

\texttt{<feedback>}your feedback\texttt{</feedback>}

\texttt{<score>}[0--2]\texttt{</score>}
}
\end{tcolorbox}
\end{center}

\paragraph{Proposal revision.}
\label{app:revision_prompt}

After receiving the critic's feedback, the proposal agent uses the following revision prompt (Prompt~7) to generate a completely new analysis from scratch, rather than editing or extending the previous draft.

\begin{center}
\begin{tcolorbox}[colback=cyan!5!white,colframe=cyan!50!black,width=0.95\linewidth,label={box:revision_prompt},title={\textit{\textbf{Prompt 7: Proposal Revision Prompt}}}]
{\small
Re-answer the original question from scratch. Using the reviewer's feedback below as reference, write a completely new analysis instead of editing or extending your previous one. Use the same output format.

Feedback: \{feedback\}
}
\end{tcolorbox}
\end{center}

\begin{figure}[htbp]
    \centering
    \includegraphics[width=0.95\linewidth]{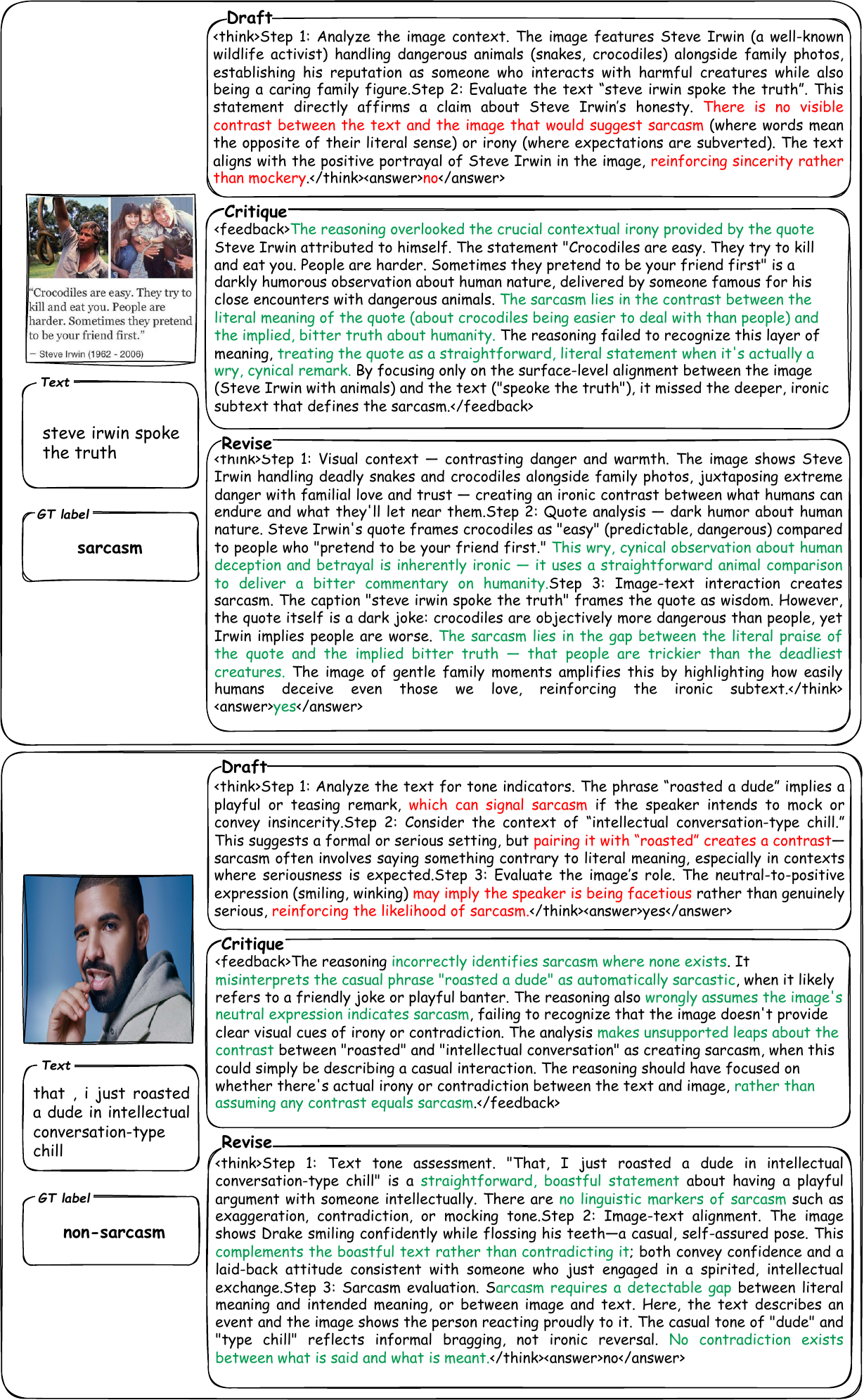}
    \caption{Successful answer correction via critic-guided revision. The draft produces an incorrect sarcasm judgment due to flawed cross-modal reasoning. The critic pinpoints the reasoning deficiency and provides targeted feedback, enabling the proposal agent to generate a corrected reasoning from scratch.}
    \label{fig:wrong_to_right}
\end{figure}

\begin{figure}[htbp]
    \centering
    \includegraphics[width=1\linewidth]{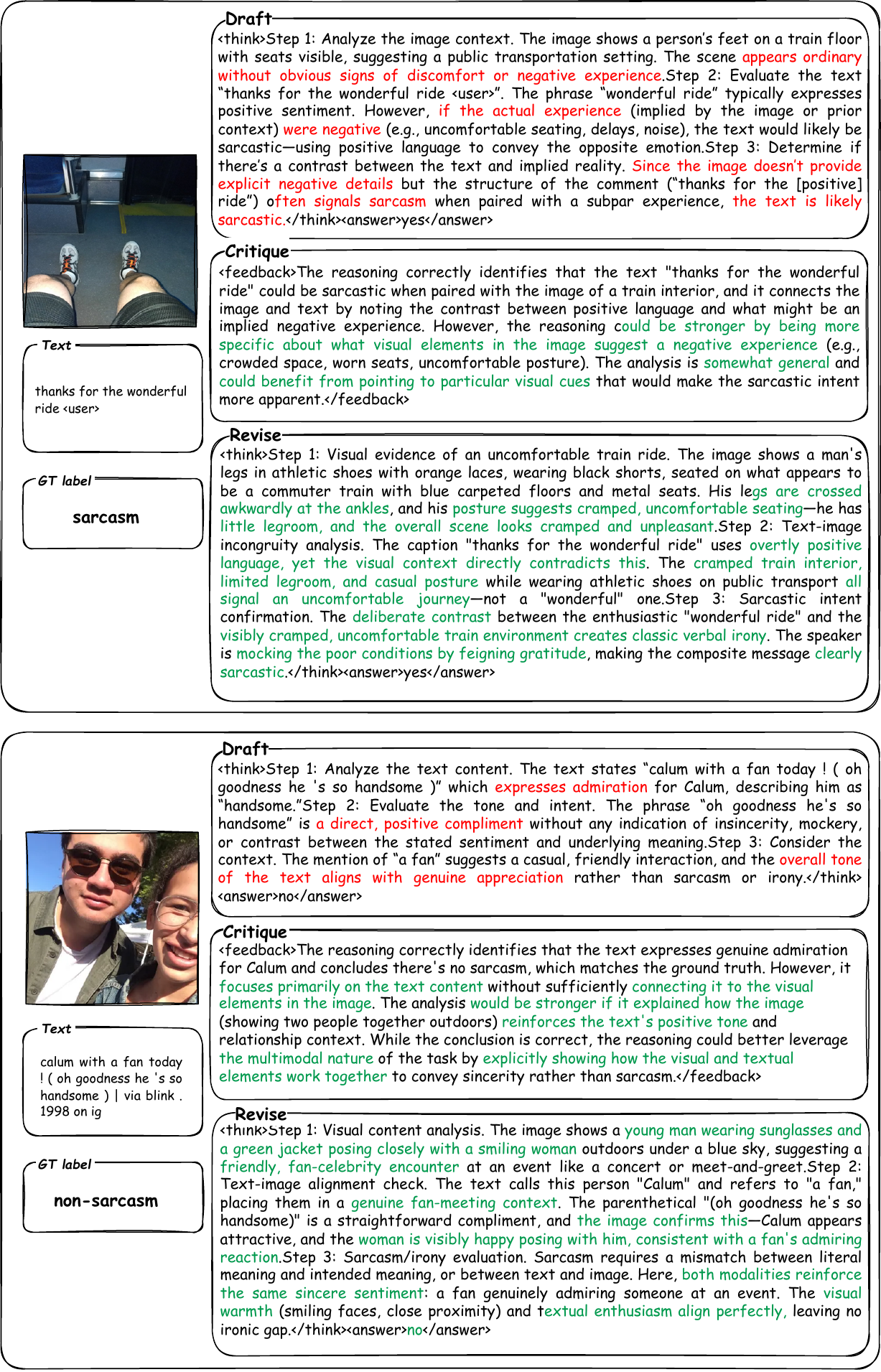}
    \caption{Reasoning quality improvement via critic feedback. The draft predicts the correct label but provides shallow, surface-level reasoning. After receiving the critic's feedback identifying analytical gaps, the revision produces a substantially deeper analysis that explicitly captures the cross-modal sarcastic mechanism.}
    \label{fig:poor_quality}
\end{figure}
%% ================================================================
%%  APPENDIX C — Additional Experiments
%% ================================================================
\section{Additional Experiments}
\label{app:additional_experiments}

This section collects supplementary experimental results and analyses that complement the main paper.

\subsection{Additional Ablations}
\label{app:additional_ablations}

% TODO: Add ablations that did not fit in the main paper, such as rollout trajectory count,
% teacher model choice, critic score granularity, and revision iteration count.

\paragraph{Effect of Multiple Revision Rounds.}
We investigate whether iterating the critique--revise cycle beyond a single round yields further improvements. Starting from the same draft, the proposal agent receives feedback from the critic agent and revises its reasoning; this process is repeated for up to five rounds on MMSD2.0. Table~\ref{tab:revision_rounds} reports the results.

\begin{table}[h]
\centering
\caption{Effect of multiple revision rounds on MMSD2.0. Each round applies one critique--revise cycle using the same critic agent.}
\label{tab:revision_rounds}
\begin{tabular}{l cccc}
\toprule
\textbf{Round} & \textbf{F1} & \textbf{Acc} & \textbf{P} & \textbf{R} \\
\midrule
Draft & 81.1 & 82.5 & 75.7 & 87.4 \\
\addlinespace
Revise 1 & {83.1} & 84.0 & 76.1 & {91.5} \\
Revise 2 & 83.1 & 84.1 & 76.6 & 90.7 \\
Revise 3 & 82.8 & 83.8 & 76.2 & 90.7 \\
Revise 4 & 83.1 & 84.1 & 76.4 & 91.0 \\
Revise 5 & 83.3 & {84.2} & {76.6} & 91.1 \\
\bottomrule
\end{tabular}
\end{table}

The first revision round contributes the dominant improvement, raising F1 from 81.1 to 83.1 (+2.0 points) with a substantial recall gain from 87.4 to 91.5.
Subsequent rounds yield marginal and inconsistent changes: F1 fluctuates within a narrow range of 82.8--83.3 across rounds 2--5, without a steady improvement.
This indicates that a single critique--revise cycle is sufficient to capture the actionable feedback from the critic agent, and additional iterations provide limited additional gains.
In practice, ProCrit therefore adopts a single revision round, which achieves the majority of the performance gain while keeping inference cost manageable.

\subsection{Efficiency Analysis}
\label{app:efficiency}
% Report training cost, inference latency, token length, and the overhead introduced by critic-guided revision.

We analyze the computational cost of ProCrit from both training and inference perspectives.

\paragraph{Training cost.}
ProCrit's training pipeline involves four stages, each operating on a 7B-parameter backbone (Qwen2.5-VL-7B) with 8 NVIDIA A800 GPUs under DeepSpeed ZeRO-3.
Table~\ref{tab:training_cost} summarizes the data scale and approximate GPU hours for each stage.
The total training budget is approximately 20 GPU hours.
Notably, all stages use modest data scales: the largest being critic SFT with 20K examples and the RL stages operate on only 2K and 5K instances, respectively.
This is substantially smaller than the hundreds of thousands of samples typically used in general-purpose reasoning RL~\cite{deepseek2025r1}, reflecting the efficiency of targeted, task-specific optimization.

\begin{table}[h]
\centering
\caption{Training cost breakdown across the full pipeline on 8$\times$A800 GPUs.}
\label{tab:training_cost}
\small
\begin{tabular}{lccc}
\toprule
\textbf{Stage} & \textbf{Data} & \textbf{Epochs} & \textbf{GPU Hours} \\
\midrule
Proposal SFT & 9K & 3 & {0.70} \\
Critic SFT & 20K & 3 & {1.80} \\
Critic GRPO & 5K & 4 & {19.50} \\
Proposal GRPO & 2K & 4 & {20.80} \\
\midrule
\textbf{Total} & & & 42.8 \\
\bottomrule
\end{tabular}
\end{table}

\paragraph{Inference cost.}
At test time, ProCrit offers two operating modes with different cost--performance trade-offs.
In \emph{draft-only} mode (ProCrit Draft), the proposal agent performs a single forward pass, incurring the same cost as any single-pass reasoning method while already achieving state-of-the-art performance (e.g., 81.1 F1 on MMSD2.0).
In \emph{draft--critique--revise} mode (ProCrit Revise), three sequential passes are required: the proposal agent generates a draft, the critic agent evaluates it and provides feedback, and the proposal agent produces a revised analysis.
This introduces additional latency relative to single-pass methods; however, we note that existing multi-agent sarcasm detection approaches such as S3Agent (3 independent reasoning agents with a decision agent) and Commander-GPT (6 sub-task agents plus a decision module) also require multiple inference calls per sample.
Importantly, a single revision round captures the dominant improvement (Table~\ref{tab:revision_rounds}), and no iterative looping is needed---the overhead is bounded and predictable.

\paragraph{Generation length.}
The proposal agent's maximum generation length is 512 tokens and the critic agent's is 768 tokens (Table~\ref{tab:hyper_grpo}).
In practice, the average draft length is considerably shorter: the majority of reasoning trajectories comprise 3 analytical steps (Table~\ref{tab:step_distribution}), with each step producing a concise, evidence-focused analysis.
The revision pass generates output of comparable length to the draft, as it produces a new analysis from scratch rather than appending to the original.
The critic's output consists of a quality score and a short natural-language feedback paragraph, typically much shorter than the 768-token budget.

%% ================================================================
%%  APPENDIX D — Qualitative Analysis
%% ================================================================

\section{Qualitative Analysis}
\label{app:qualitative}

% Figure~\ref{fig:case_study} illustrates representative examples from the MMSD2.0 test set.

This section presents additional examples to illustrate the behavior of the draft--critique--revise pipeline on concrete samples from the MMSD2.0 test set.
As shown in Figure~\ref{fig:wrong_to_right}, the critic agent is able to identify specific reasoning deficiencies in incorrect drafts---such as misinterpreted pragmatic cues or overlooked contextual irony---and provide targeted feedback that guides the proposal agent toward a corrected analysis generated from scratch.
As demonstrated in Figure~\ref{fig:poor_quality}, the critic also contributes when the draft already arrives at the correct prediction but with insufficiently grounded reasoning: it pinpoints analytical gaps, prompting the proposal agent to produce a substantially deeper revision that explicitly anchors its judgment in concrete multimodal evidence.
These cases highlight that the critic serves a dual role---not only as an error-correction mechanism but also as a quality-enhancement signal that strengthens the analytical rigor of the reasoning process.

% \begin{figure}
%     \centering
%     \includegraphics[width=1\linewidth]{figures/failed_cases.pdf}
%     \caption{Failed self-correction caused by over-analysis. The model's draft initially arrives at the correct label through robust reasoning. Upon receiving the critic agent's feedback questioning its initial assumptions, the revision second-guesses the correct rationale, resulting in overthinking that corrupts the final output and predicts the wrong label.}
%     \label{fig:failed_cases}
% \end{figure}

%% ================================================================
%%  APPENDIX E — Limitations
%% ================================================================
\section{Limitations}
\label{app:limitations}

% TODO: Expand with more specific discussion.

% The current evaluation is conducted on English-language social media datasets (Twitter and Reddit). Sarcastic expression varies considerably across languages and cultural contexts, making multilingual and cross-platform validation a valuable direction. Since ProCrit's core components do not encode language-specific or domain-specific assumptions, such extension is readily achievable by re-synthesizing training annotations from the target domain.
% In addition, the current formulation targets binary sarcasm detection. Extending the multi-perspective reasoning paradigm to finer-grained tasks---such as sarcasm type classification and target identification---would broaden its applicability and is naturally supported by the framework's flexible analytical structure.
% Finally, sarcasm perception is inherently subjective---the same expression may be interpreted as sarcastic by some audiences but not others, depending on shared context, cultural background, and interpersonal familiarity.
% The current benchmarks adopt majority-vote labels that inevitably smooth over such disagreements.
% Incorporating richer contextual signals (e.g., conversational history or author profile) and modeling annotator disagreement as soft labels could better capture this subjectivity and further improve robustness.
The draft--critique--revise paradigm requires three sequential 
forward passes at inference time (proposal draft, critic 
evaluation, and proposal revision), increasing computational 
cost compared to single-pass methods. 
We provide a detailed efficiency analysis in 
Appendix~\ref{app:efficiency} and note that the draft-only 
output already achieves competitive performance, offering a 
practical alternative when inference cost is a concern.
Moreover, the current evaluation is conducted on English-language social media datasets (Twitter and Reddit). Sarcastic expression varies considerably across languages and cultural contexts, and validating the framework in multilingual and cross-platform settings remains an important direction for future work.
% In addition, the current formulation targets binary sarcasm detection. Extending the multi-perspective reasoning paradigm to finer-grained tasks---such as sarcasm type classification, intent interpretation, or target identification---would broaden its applicability and is naturally supported by the framework's flexible analytical structure.
Additionally, sarcasm perception is inherently subjective---the same expression may be interpreted as sarcastic by some audiences but not others, depending on shared context, cultural background, and interpersonal familiarity.
The current benchmarks adopt majority-vote labels that inevitably smooth over such disagreements.
Incorporating richer contextual signals (e.g., conversational history or author profile) and modeling annotator disagreement as soft labels could better capture this subjectivity and further improve robustness.

%% ================================================================
%%  APPENDIX F — Broader Impact
%% ================================================================
\section{Broader Impact}
\label{app:broader_impact}

% ProCrit advances multimodal sarcasm detection by producing explicit, multi-perspective reasoning alongside predictions, offering greater interpretability than end-to-end classifiers.
% This transparency has practical value in content moderation and social media analysis, where understanding \emph{why} a system flags content as sarcastic is as important as the prediction itself---human reviewers can inspect the reasoning process and make informed decisions rather than acting on opaque outputs.

% We note that sarcasm detection, like all NLU tasks, should be deployed as a decision-support tool rather than a fully autonomous system.
% Sarcastic intent is inherently ambiguous and context-dependent, and any automated system---including ProCrit---should be complemented by human judgment in high-stakes applications.
% Additionally, the training data is drawn from English-language social media, and practitioners should evaluate domain suitability before deployment in other linguistic or cultural settings.

% ProCrit is designed to assist content moderation and social media analysis by providing explicit, multi-perspective reasoning alongside sarcasm predictions, enabling human reviewers to make more informed decisions.
% As with any NLU system, outputs should be treated as references for human judgment rather than definitive conclusions, particularly given the inherent subjectivity of sarcastic expression across cultural and linguistic contexts.

Failure to recognize sarcasm can affect the accurate interpretation of sentiment and communicative intent in applications such as opinion mining, public sentiment analysis, and online content moderation.
By improving the accuracy and robustness of multimodal sarcasm detection, ProCrit may help such systems better capture the intended meaning behind social media posts.
In addition, the reasoning chains generated alongside predictions can provide greater transparency into the decision-making process, enabling more effective human oversight in sensitive deployment scenarios.
We also acknowledge potential risks. Sarcasm detection systems 
could be misused for surveillance or censorship of online speech. 
Moreover, since sarcastic expression is culturally situated, 
deploying a model trained predominantly on English-language 
social media in other cultural contexts may produce systematic 
errors. We therefore recommend that ProCrit be used as a 
decision-support tool complementing human judgment rather than 
as a fully autonomous content filter.

%%%%%%%%%%%%%%%%%%%%%%%%%%%%%%%%%%%%%%%%%%%%%%%%%%%%%%%%%%%%

\end{document}